\documentclass[iop,apj,tighten]{emulateapj}
\usepackage[utf8x]{inputenc}
\usepackage{graphicx}
\usepackage{textcomp}
\usepackage{bm}
\usepackage{amsmath}
\slugcomment{Accepted for publication in The Astrophysical Journal.}
\shorttitle{Magnetic Fields and Acoustic Waves}
\shortauthors{Rajaguru {\it et al.}}

\begin{document}
\title{Magnetic fields and low-frequency acoustic wave-energy supply to the solar chromosphere}
\author{S.P. Rajaguru}
\affil{Indian Institute of Astrophysics, Bangalore-34, India}
\email{rajaguru@iiap.res.in}

\author{C.R. Sangeetha} 
\affil{Inter-University Centre for Astronomy and Astrophysics, Post Bag-4, Ganeshkhind, Pune 411007, India}
\email{sangeetha@iucaa.in}

\author{Durgesh Tripathi}
\affil{Inter-University Centre for Astronomy and Astrophysics, Post Bag-4, Ganeshkhind, Pune 411007, India}
\email{durgesh@iucaa.in}

\begin{abstract} 
The problem of solar chromospheric heating remains a challenging one with wider implications for stellar physics. 
Several studies in the recent past have shown that small-scale inclined magnetic field elements channel copious amount of energetic 
low-frequency acoustic waves, that are normally trapped below the photosphere. These magneto-acoustic waves are expected to shock at 
chromospheric heights contributing to chromospheric heating. In this work, exploiting simultaneous photospheric vector magnetic field, 
Doppler, continuum and line-core intensity (of FeI 6173 {\AA}) observations from the Helioseismic and Magnetic Imager (HMI) and 
lower-atmospheric UV emission maps in the 1700 {\AA} and 1600 {\AA} channels of the Atmospheric Imaging Assembly (AIA), both onboard 
the Solar Dynamics Observatory (SDO) of NASA, we revisit the relationships between magnetic field properties (inclination and strength) 
and the acoustic wave propagation (phase travel time). We find that the flux of acoustic energy, in the 2 - 5 mHz frequency range, 
between the upper photosphere and lower chromosphere is in the range of 2.25 - 2.6 kW m$^{-2}$, which is about twice the previous 
estimates. We identify that the relatively less-inclined magnetic field elements in the quiet-Sun channel a significant amount of waves 
of frequency lower than the theoretical minimum for acoustic cut-off frequency due to magnetic inclination. We also derive indications 
that these waves steepen and start to dissipate within the heights ranges probed, while those let out due to inclined magnetic fields 
pass through. We explore connections with existing theoretical and numerical results that could explain the origin of these waves.

\end{abstract} 


\section{Introduction}\label{sec:intro}

Identifying and understanding the mechanisms of non-thermal supply of energy to the 
atmospheric layers (chromosphere and corona) of stars with outer convection zones so as to fully account for the observed
radiative losses continue to remain a major problem in astrophysics \citep{1996SSRv...75..453N,2006SoPh..234...41K,2017ARA&A..55..159L}. 
There have been enormous amount of work and progress in this field, and the proposed mechanisms can be grouped under two major theories:
(i) mechanical heating by waves, both non-magnetic \citep{1946NW.....33..118B,1948ApJ...107....1S,1990ASPC....9....3U} and magnetic
\citep{1947MNRAS.107..211A,1996SSRv...75..453N, 2003ASPC..286..363U}, that are generated 
by turbulent convection in the photosphere and below \citep{1952RSPSA.211..564L,1967SoPh....2..385S,1968ApJ...154..297S} 
or by a host of magneto-hydrodynamic wave processes that couple the convection and acoustic waves to magnetic fields \citep{1961ApJ...134..347O,
1997ApJ...486L..67C,2002ApJ...564..508R,2003ApJ...599..626B,2013JPhCS.440a2048K}, and (ii) direct dissipation of magnetic energy through 
Joule heating, {\it i.e.,} resistive dissipation of electric currents, in sheared magnetic configurations and during magnetic reconnections 
(flares, micro- and nano-flares) \citep[e.g.,][]{1991ApJ...376..355P} with associated high speed flows and jets \citep[see e.g.,][and references therein]
{2004Natur.430..536D,2013ApJ...770L...3K,2015MNRAS.446.3741C,2013ApJ...778..142T}. 
Although the chromosphere acts as gateway to overlying hotter layers, where the above second group of processes are observed dominantly, 
much of the progress in the field has favoured the wave-heating mechanisms (under the first of above two theories) as the dominant ones 
in it \citep{2005ApJ...633L..57S,2006ApJ...648L.151J,2007ApJ...671.2154K,2008ApJ...681L.125S}. This also has some bearing on the
much larger observational characterisation of stellar chromospheres through the emission in the cores of H and K lines of Ca II 
\citep{1984ApJ...279..763N,2007ApJ...671.2154K}: acoustic heating is thought to provide the so called 'basal flux' of emission in the above
chromospheric lines in Sun-like low emission slow rotators \citep{1989ApJ...341.1035S}. However, the exact nature of waves that
heat the chromosphere has been brought into debate after \citet{2005Natur.435..919F} estimated that the flux of acoustic waves that reach
the quiet-chromosphere (apparently non-magnetic and hence acoustic waves of frequency larger than the quiet-Sun cut-off of 5.2 mHz)
is too small to balance the radiative losses. 
 
While the exact amount of energy flux in high-frequency acoustic waves and the dependence of its estimate on spatial resolution 
are still being debated \citep{2007ApJ...671.2154K,2009A&A...508..941B,2010A&A...522A..31B}, the \citet{2005Natur.435..919F} result 
also brought into question the 'non-magnetic chromosphere' with the possibility that waves modified by the magnetic field might supply
the required flux of energy. In this connection, the findings of \citet{2006ApJ...648L.151J} that the small-scale inclined magnetic 
field elements comprising the quiet-network act as magneto-acoustic portals through which copious amount of low-frequency (2 - 5 mHz)
acoustic waves propagate outward extracting energy from the acoustic reservoir below acquire prominence.
Normally, acoustic waves of frequency less than the photospheric cut-off of 5.2 mHz, estimated from $\omega_{ac} = c/2H_{\rho}$, 
where $c$ and $H_{\rho}$ are the photospheric sound speed and density scale-height, respectively, are trapped below the photosphere 
and resonate as {\it p} modes. In the presence of magnetic fields, simple theoretical considerations \citep{1977A&A....55..239B} show that, 
when the plasma $\beta$ is low ($\beta < $1), the $\omega_{ac}$ is reduced by a factor of $cos(\theta)$, where $\theta$ is the inclination 
of magnetic field with respect to the direction of gravity (local normal to the solar surface). Such effects of magnetic field have now been 
well attested as being responsible in a varied set of phenomena \citep{2004Natur.430..536D, 2005ApJ...624L..61D, 2006ApJ...647L..77M, 
2007ApJ...654L.175R, 2010ApJ...721L..86R}; see also \citet[][and references therin]{2016GMS...216..489C}. 

The previous study of energy flux in low-frequency waves in the small-scale quiet-network \citep{2006ApJ...648L.151J}, 
propagating out due to reduction in acoustic cut-off frequency, used magnetic inclination angle derived from potential field 
extrapolation of line-of-sight (LOS) magnetic field observed by SoHO/MDI \citep[MDI;][]{1995SoPh..162..129S} with the wave-data derived from 
Doppler velocities observed by the South Pole instrument Magneto-Optical filter at Two Heights experiment (MOTH) \citep{2004SoPh..220..317F}. 
Similarly, the study of magnetic reduction of acoustic cut-off over a sunspot region \citep{2006ApJ...647L..77M} using 1600~{\AA} intensities 
from the Transition Region and Coronal Explorer (TRACE) \citep{1999SoPh..187..229H} too utilised magnetic inclination derived from the LOS B from SoHO/MDI.
In this work, exploiting simultaneous photospheric vector magnetic field, Doppler, continuum and line-core intensity (of FeI 6173~{\AA})
observations from the Helioseismic and Magnetic Imager (HMI) and the two lower-atmospheric UV emission maps
in the 1700~{\AA} and 1600~{\AA} wavelength channels of the Atmospheric Imaging Assembly (AIA), both onboard the
Solar Dynamics Observatory (SDO) of NASA,  we revisit the relationships between magnetic field properties (inclination and strength)
and the acoustic wave propagation (phase travel time). The above two instruments on the same platform provide simultaneous
multi-height observations to observe the waves (through velocities and intensities) as well as the vector magnetic fields,
and with higher spatial resolution than the above referred previous such studies.
 
In addition to our primary aim of studying (magneto-)acoustic energy supply to chromosphere via the small-scale magnetic network, 
in order to check and compare such processes in a range of background magnetic fluxes or sizes of structures,
we include plage and spot regions too (see, e.g. review by \citet{2013JPhCS.440a2048K}). In particular, including quiet and active Sun observations from
the above described space-platform, we are able to examine closely the relationship between magnetic inclination, acoustic cut-off and
height in the atmosphere, and derive acoustic wave energy flux as a function of magnetic field incliantion
and strength to connect with recent theoretical studies \citep{2006MNRAS.372..551S,2013MNRAS.435.2589C}.
Rest of the paper is structured as follows: in
\S\ref{dam} we describe the data and analysis methods used in this study. In \S\ref{main} we present and discuss the results
in several sub-sections followed by a summary and conclusion in \S\ref{main}.

\section{Data and Analysis Method}\label{dam}
\subsection{Data}\label{dat}
The HMI instrument uses the magnetically sensitive line Fe {\sc i} 6173.34 \AA~ for Doppler and vector magnetic field observations 
\citep{2012SoPh..275..207S}. The observables derived are photospheric continuum intensity $I_{\rm{c}}$ near this line, line depth $L_{\rm{d}}$, 
Doppler velocity $v_{\rm{d}}$, and magnetic field, both line-of sight $B_{\rm{los}}$ and vector. The cadence of $I_{\rm{c}}$, $L_{\rm{d}}$, 
$v_{\rm{d}}$ and $B_{\rm{los}}$ is 45~s. Vector magnetic field has a slower cadence of 720~s as standard pipe-line product, but can be 
obtained at a cadence of 135~s too \citep{2014SoPh..289.3483H}. The AIA observes the Sun using nine wavelength channels, namely 1700, 1600, 
304, 171, 193, 211, 335, 94, \& 131~{\AA} covering different layers of the solar atmosphere \citep{2012SoPh..275...17L}. For our purpose here, 
AIA observations taken in 1700~{\AA} and 1600~{\AA} channels that map the lower solar atmosphere are sufficient. 
These images are captured with a cadence of 24~s. HMI and AIA instruments observe the Sun's disk with a pixel size of $\sim$~0.5~\arcsec 
and 0.6~\arcsec, respectively. Hereafter, we shall refer to the observed intensities at 1700~{\AA} as $I_{\rm{uv1}}$ and that at 1600~{\AA} as $I_{\rm{uv2}}$. 
Since $I_{\rm{c}}$ is dominated by granulation (see discussion in \S\ref{phcoh}), we add line core intensity determined as 
$I_{\rm{co}} = I_{\rm{c}} - L_{\rm{d}}$ to the HMI observables used in our studies.
\begin{figure*}[ht]  
\centering
\epsscale{0.95}
\plottwo{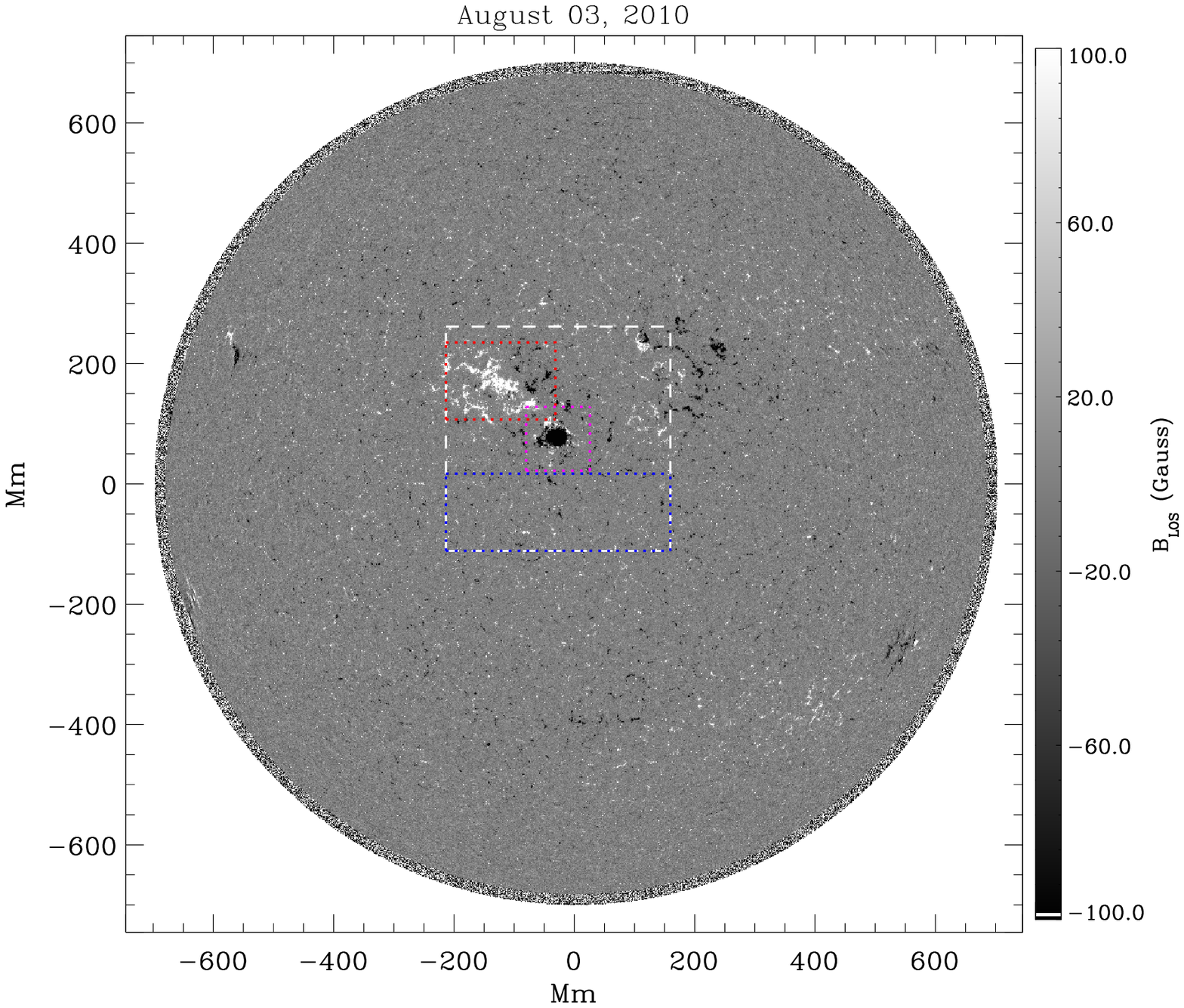}{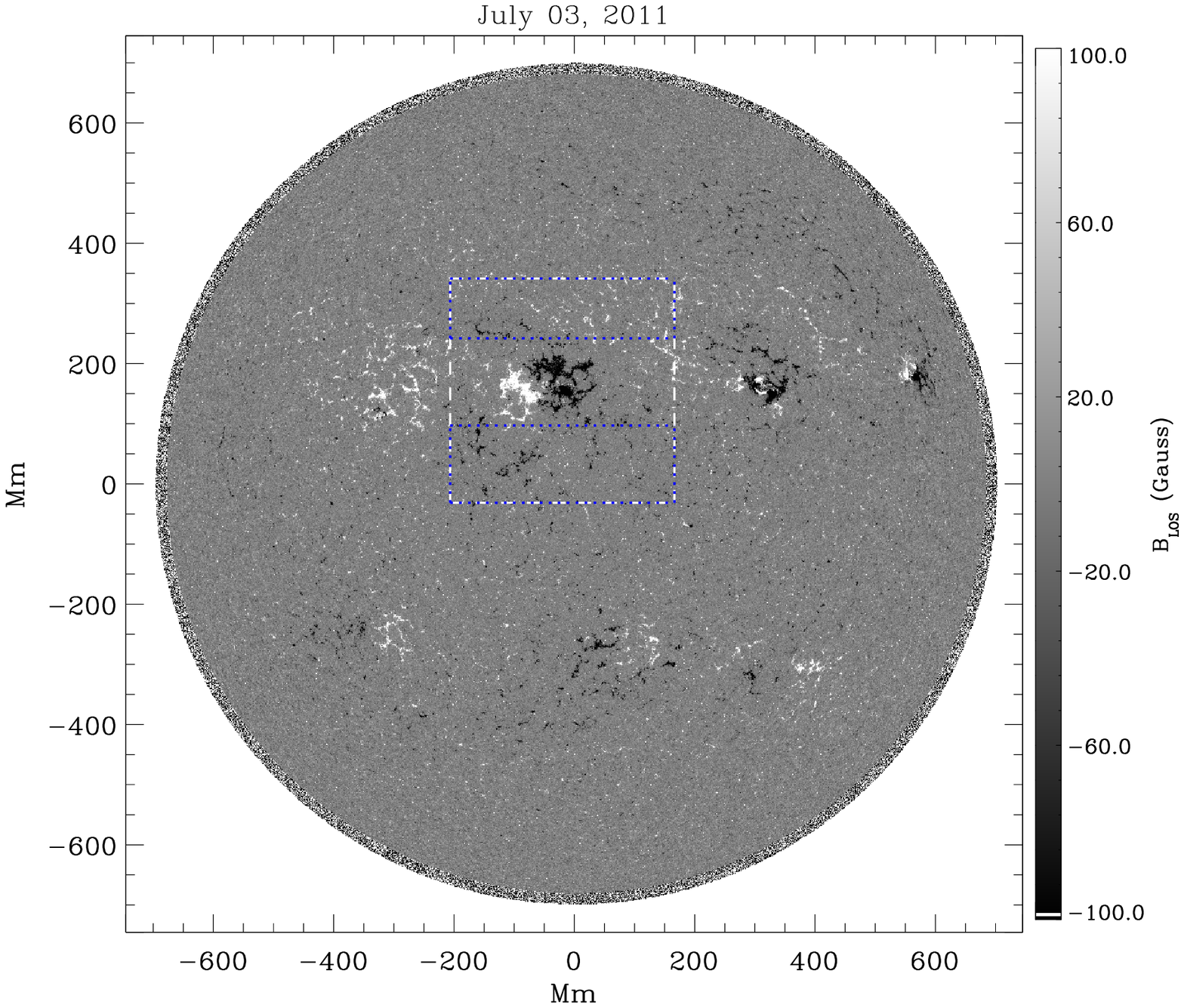}
\caption{Context images showing the full-disk line-of-sight magnetic field, $B_{LOS}$, corresponding to the mid-time of observation
period used in this work. The two images correspond to the two central meridian crossing dates as marked on the top of panels.
Two large square regions on each date, marked by white long-dashed boundaries, are tracked in all observables used in this work.
The coloured dotted-line boundaries mark the sub-regions studied in this work: regions enclosed by blue, red and pink dotted lines
are the quiet, plage and spot regions, respectively. The grey-scale of the image has been satured at $\pm$~100~G to make the small-scale
magnetic fields clearly visible.}\label{fig1}
\end{figure*}

For the study here, we require all the above HMI and AIA observables, used to trace the waves, co-aligned and tracked with same spatial resolution. 
The JSOC ({\it http://jsoc.stanford.edu }) helioseismology pipeline facilitate this and we use it.
Here, after tracking, the AIA data $I_{uv1}$ and $I_{uv2}$ are resampled to match the HMI cadence of 45~sec.  
In an earlier study of high-frequency acoustic halos \citep{2013SoPh..287..107R} such data were prepared
for four large regions each covering a sunspot and quiet-Sun areas.
Here, we choose two regions among them with a larger fraction of quiet-Sun network with central meridian 
crossing dates of 03 August 2010 and 03 July 2011. Fig.~\ref{fig1} displays context images of the whole Sun in LOS magnetic field corresponding to mid-time of
observation period used, and the two regions studied are marked with white dashed boundaries.
Note that we have displayed the magnetograms with the gray-scale saturated beyond $\pm$~100~G so as to make the small-scale magnetic network clearly visible.
The coloured boundaries (dotted lines) mark the quiet-network (blue), plage (red)  and spot (pink) sub-regions.  
Both the acoustic waves, which are excited by turbulent convection, and the quiet-Sun magnetic elements are highly dynamic 
and exhibit stochastic behaviour. Hence, the interaction between them, here especially the magnetic inclination caused
propagation of acoustic waves that we want to track, is intermittent and stochastic \citep{2006ApJ...648L.151J}; a suitable
compromise for the total length of data over which the signatures of such interactions, here wave travel-times between heights
and power spectrum of waves, can be averaged with sufficient signal-to-noise ratio is to be chosen. Here, for
the data that we use, 14~hours is found sufficient. 

The vector magnetic field over the region of study is extracted using the HMI vector-field pipeline \citep{2014SoPh..289.3483H}.
It employs a Milne--Eddington based algorithm (the {\em Very Fast Inversion of the Stokes Vector} - VFISV: \citet{2011SoPh..273..267B}) for 
inverting the Stokes parameters and an improved version of the {\em minimum energy algorithm }\citep{1994SoPh..155..235M,2006SoPh..237..267M,
2009SoPh..260...83L} for resolving the 180\textdegree~azimuthal ambiguity in transverse field. 
We use the standard 12-minute cadence vector field maps, which, over the target region,
are then tracked and remapped the same way as the other data used in this study. The resulting products are
the disambiguated vector field maps [$B_{\rm{x}}, B_{\rm{y}}$, $B_{\rm{z}}$].
\begin{figure*}[ht]
\centering
\includegraphics[width=0.7\textwidth]{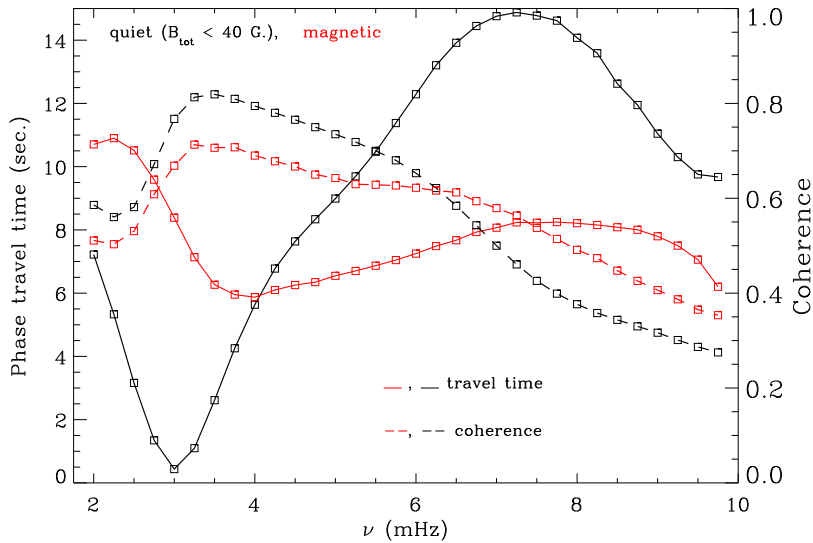}
\caption{Quiet and magnetic area averages of $t_{\rm{ph}}$ and $C$ against frequency $\nu$ from within the quiet-network regions enclosed
        by the blue dotted-line boundaries of Fig.~\ref{fig1}.}\label{fig2}
\end{figure*}

For the purpose of this study (similar to \citet{2011SoPh..268..349S,2013SoPh..287..107R}) we use a field
inclination $\xi$ {\footnote{we use $\xi$ instead of $\gamma$ because the latter is commonly used for field inclination with respect to the line-of-sight}}
defined as $\tan(\xi) = B_{\rm{z}}/B_{\rm{h}}$, where $B_{\rm{h}}=\sqrt{B_{\rm{x}}^{2}+B_{\rm{y}}^{2}}$.
Hence, $\xi$ varies between $\pm$ 90\textdegree~with $\xi$ = 0\textdegree~when the field is horizontal, 
$\xi < 0^{\circ}$ when $B_{\rm{z}} < 0$ and $\xi > 0^{\circ}$ when $B_{\rm{z}} > 0 $.
The total magnetic field strength is $B_{\rm{tot}}=\sqrt{B_{\rm{x}}^{2}+B_{\rm{y}}^{2}+B_{\rm{z}}^{2}}$.

In the presence of inclined magnetic field, as briefly discussed in \ref{sec:intro}, the acoustic cut-off frequency, $\nu_{\rm{ac}}$,
reduces depending on the local plasma $\beta$ and the inclination, $\theta$, of the magnetic field to the solar surface normal 
(direction of gravity) \citep{1977A&A....55..239B}; the thoretical lower limit is given by $\nu^{B}_{\rm{ac}}=\nu_{\rm{ac}}cos\theta$. 
The two angles $\theta$ and $\xi$, then, are related as $\theta = 90 \ - \ |\xi|$.

The different observables from HMI and AIA discussed above correspond to different heights in the photosphere and above.
The formation process of HMI line Fe {\sc i} 6173.34 \AA~ has been studied by several authors \citep{2006SoPh..239...69N,2009A&A...494.1091B,
2011SoPh..271...27F}, who derive a height range of $z$ = 20 -- 300 km above the continuum optical depth $\tau_{\rm{c}}$ = 1 ($z$=0 km) level 
with slight differences in the mean height (250 to 300 km) for the line-core formation. As to the derived observables from HMI 
[see \citet{2012SoPh..278..217C}], analyses comparing them with line synthesis of 3D simulations \citep{2011SoPh..271...27F,2014SoPh..289.3457N} show that 
$v_{\rm{d}}$ forms at a mean height of 150 km, while the HMI line core near about 230 km; the M-E inversion algorithm is most sensitive to Stokes
signals near the line-core and hence a height of 200 km is appropriate for the HMI vector magnetic field.
In a detailed analysis of the performance of observables processing algorithms for HMI, \citet{2012SoPh..278..217C} found that
while the derived $I_{\rm{c}}$ agrees well with continuum intensity near the line, line depth $L_{\rm{d}}$ is underestimated by about 30 - 33~\%. 
This would then mean that, with a height of 250 km for the actual line-core, the derived line-core intensity $I_{\rm{co}}$ corresponds to a height 
between 0.67 to 0.7 $\times$ 250 which is $\approx$ 170 km  (assuming a linear relation between height of formation and the spectral intensity within the line). 
The AIA 1700 \AA~ and 1600 \AA~ intensities [$I_{\rm{uv1}}$ and $I_{\rm{uv2}}$], form at mean heights of 360 km and 430 km \citep{2005ApJ...625..556F}, 
respectively. 

\subsection{Phase (travel-times) and Coherence Spectra}\label{phcoh}

We use cross-spectra between a pair of same physical variables corresponding to two heights to derive phase travel times, $t_{\rm{ph}}$, of waves
between heights as a function of frequency, $\nu$. Since we have velocities ($v_{\rm{d}}$) at only one height, we use intensity - intensity ($I - I$) \
cross-spectra. 
Although velocities provide less noisy capture of wave-motions, both due to p modes (helioseismic waves) that are evanescent in the photospheric 
layers \citep{1993ApJ...410..829D,2001ApJ...561..444S} and waves propagating out in the atmosphere \citep{2006ApJ...648L.151J,2009A&A...508..941B}, 
intensities have been extensively used in studies of compressive waves in the solar atmosphere \citep[e.g.,][]{2001A&A...379.1052K,2005A&A...430.1119D,
2010ApJ...721..744K} including those that derive phase shifts over height \citep[e.g.,][]{2006ApJ...647L..77M,2016ApJ...830L..17Z}. Non-adiabatic 
evolution due to radiative damping (RD, hereafter), however, will significantly modify the waves at heights where radiative relaxation time 
\citep{1957ApJ...126..202S}, $\tau_{\rm{R}}$, is comparable to or shorter than wave periods, $\tau_{\rm{S}}$ = 1/$\nu$ \citep{1972A&A....17..458S,
1978SoPh...57..245S,1980SoPh...68...87W,1990A&A...236..509D,2011MNRAS.417.1162N}. Near the transition $\tau_{\rm{R}} \approx \tau_{\rm{S}}$ RD can potentially 
be different in velocities and intensities and hence may contribute to the $I - V$ cross-spectra \footnote{The usually observed signals 
in $I - V$ cross-spectra pertain to a given height or spectral line: the evanescent p modes (the 2 - 5 mHz range ridges in the $k_{\rm{h}} - \nu$ diagram)
exhibit larger than the expected adiabatic phase shifts of 90\textdegree, while the inter-ridge background region with $\nu$ less than about 
3.0 mHz shows negative or near-zero phase-shifts \citep{1990A&A...236..509D,2001ApJ...561..444S,2013SoPh..284..297S}. 
Although a complete physical model capturing all the features in $I - V$ spectra is still lacking, existing literature 
includes successful models based on the correlated noise background (turbulent convection that excites the oscillations) to explain the inter-ridge
background and modal asymmetries in the $I$ and $V$ power spectra \citep{1999ApJ...510L.149N,2000ApJ...535..464S} while other models that 
include non-adiabaticity due to radiative damping claim better match with observations \citep{2013SoPh..284..297S}.}
\citep{1990A&A...236..509D,1991sia..book..329H,2013SoPh..284..297S}; in
general, when $\tau_{\rm{R}} > \tau_{\rm{S}}$ or comparable, the acoustic cut-off frequency is not altered much and hence the character of waves, 
evanescent or propagating, is retained. When $\tau_{\rm{R}} << \tau_{\rm{S}}$, apart from radiatively damping the already propagating
waves \citep{1972A&A....17..458S,2011MNRAS.417.1162N} RD can also cause the evanescent waves to propagate by significantly lowering the 
acoustic cut-off frequency \citep{1978SoPh...59...49G,1983SoPh...87...77R,2006RSPTA.364..447R,2006ASPC..358..465C,2008ApJ...676L..85K} -- 
we discuss this further in \S\ref{res2.ph} and \S\ref{sum} in relation to the results obtained in this work. 

In the study here, however, RD {\em per se} does not contaminate phase shifts between heights but its height-variation, due to that in 
$\tau_{\rm{R}}$, can. Height variation of $\tau_{\rm{R}}$ closely follows that of temperature and exhibits strong gradients close to the 
$\tau$ = 1 photospheric region and higher up in the transition layer and corona \citep{1978SoPh...57..245S,1982ApJ...263..386M,2013SoPh..284..297S}. 
In the intervening upper photospheric - lower chromospheric region, $\tau_{\rm{R}}$ is nearly constant:
between $\approx$ 170 - 200 km (HMI line core intensity $I_{\rm{co}}$) and $\approx$ 360 - 430 km (AIA 1700~{\AA} and 1600~{\AA})
that the results in this work correspond to, we note from the realistic estimates given in Figure 4 (upper panel) of \citet{2013SoPh..284..297S} that
$\tau_{\rm{R}}$ is nearly constant in the range of 30 - 50 secs as compared to the wave periods of interest (180 - 500 secs). 
This is reflected in the nearly constant $I - V$ phase difference, between 200 - 500 km, shown in Figure 4 (lower panel) of \citet{2013SoPh..284..297S}
for a 3.3 mHz p mode and also confirms to the results of \citet{1985PhDT.......213S} (see also \citet{1984MmSAI..55..147S} and a discussion
by \citet{1991sia..book..329H}) shown in Figure 1 of \citet{1990A&A...236..509D}. Hence artificial phase signals due to height-dependent RD 
in the $I(z_{1}) - I(z_{2})$ or $v(z_{1}) - v(z_{2})$ phase differences between heights $z_{1}$ and $z_{2}$ in the above range should be negligible.
Note that the $I - V$ phase difference itself is not the expected 90\textdegree~for adiabatic evanescent p mode but it is near about $\approx$ 
130\textdegree~which can be due to RD \citep{2013SoPh..284..297S}, but the relevant fact here is that this value remains roughly constant over 
the height range considered here. This assessment is further substantiated in \S\ref{res1}, where we show that quiet-Sun phase shifts derived 
from intensities match very closely those from velocities over a similar height range \citep{2016ApJ...819L..23W}.

The two-height cross-spectra here are between two time-series, and are calculated at each location (pixel). Hence the derived $t_{\rm{ph}}$ are maps, \rm{i.e.,} 
$t_{\rm{ph}}(x,y,\nu)$ where $x$ and $y$ are the spatial coordinates. 
The cross-spectrum of two time-series $I_{\rm{1}}$ and $I_{\rm{2}}$ is defined as the complex-valued product of the Fourier transforms of the two series,
\begin{equation}
    \label{eq:1}
X_{12}(\nu)= {\bm I}_{\rm{1}}(\nu){\bm I^{*}}_{\rm{2}}(\nu)
\end{equation}
where ${\bm I}$'s are the Fourier transforms; $*$ represents the complex conjugate. The phase-lag, $\delta\phi$, between the two time-series is then
given by the phase of the complex cross-spectrum $X_{12}(\nu)$,
\begin{equation}
    \label{eq:2}
\delta\phi(\nu) = \tan^{-1}[Im(<X_{12}(\nu)>)/Re(<X_{12}(\nu)>)]
\end{equation}
The magnitude of $X_{12}(\nu)$ is used to calculate the coherence,
\begin{equation}
    \label{eq:3}
C(\nu) = \frac{|<X_{12}(\nu)>|^{2}}{<|{\bm I}_{\rm{1}}|^{2}><|{\bm I}_{\rm{2}}|^{2}>},
\end{equation}
which ranges from 0 to 1 and is a measure of linear correlation between $I_{\rm{1}}$ and $I_{\rm{2}}$.
In the above two equations, $<.>$ denotes ensemble average or expectation value, which requires averaging over 
realisations. In our case here, each realisation corresponds to a wave-process in which an acoustic wave, recorded in $I_{\rm{1}}$,
hits the magnetic field, which depending on its inclination to solar normal (and to the wave-vector)
lets it propagate outward to a height where it is recorded as $I_{\rm{2}}$. For a single realisation of a wave
connecting the two heights, coherence at a frequency is unity. Normally, a frequency averaging is performed
to estimate coherence. Given a granular lifetime of ~10 minutes, and that we are mainly interested in acoustic waves of periods
in the range of 3 - 10 minutes (1.5 - 5.2 mHz), we do have a large number of independent realisations of the above
wave-process in the 14-hour long observations that we use. Hence, to achieve the above ensemble averaging, we split the
time-series into segments of equal length of about ~ 96 mins (corresponding to 128 data points in the 45 sec. cadence data)
and calculate cross-spectra for each segment and then average them together. 
We believe that this way of estimating $\delta\phi$ and $C$ does better justice to the dynamic and intermittent nature of
interactions between the magnetic field and acoustic waves. The basic cross-spectral estimation as above is implemented
using a IDL (Interactive Data Language) code by Simon Vaughan available at \url{https://github.com/svdataman/IDL}.
\begin{figure*}[ht] 
$\begin{array}{rl}
    \includegraphics[width=0.5\textwidth]{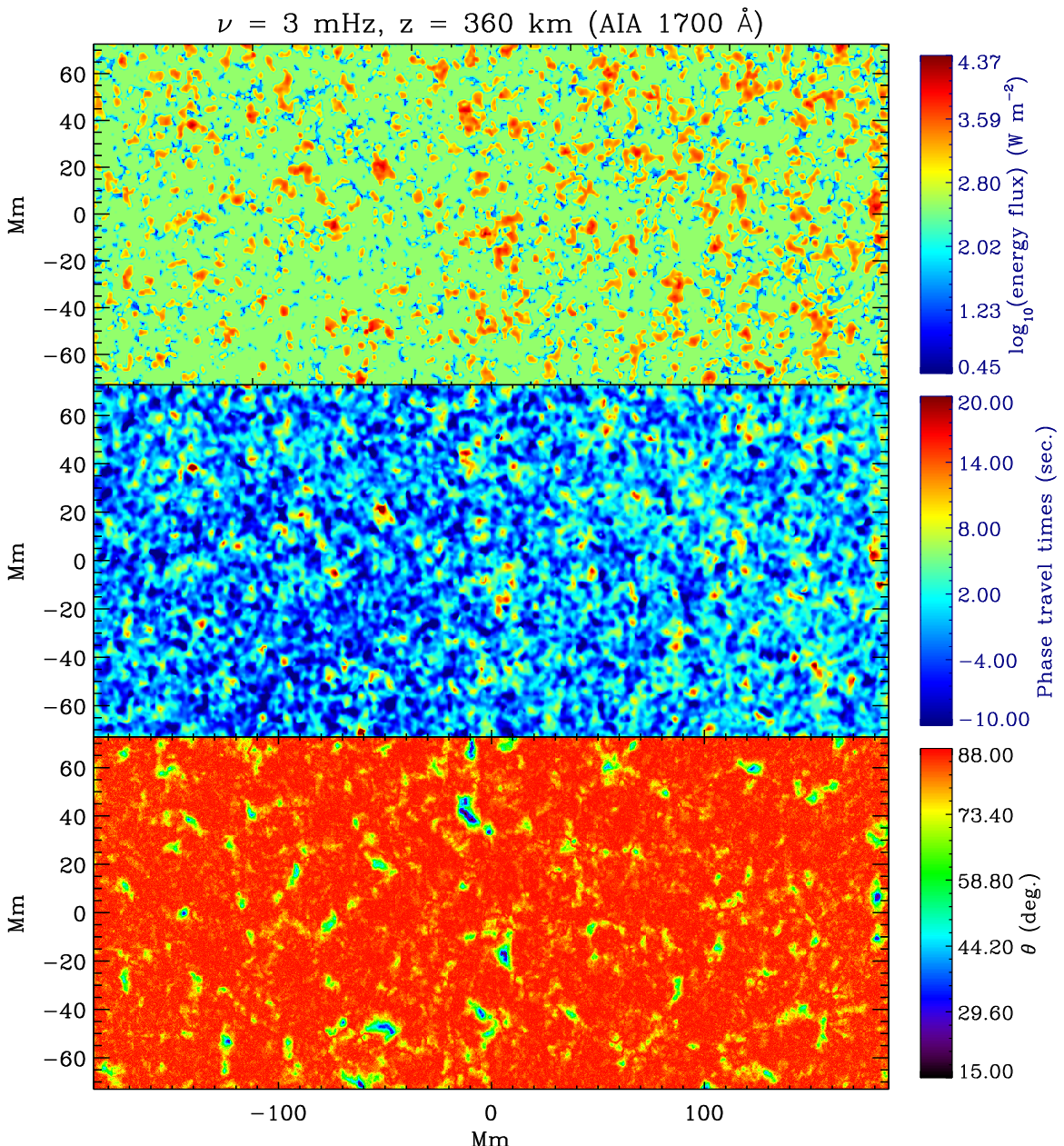} &
    \includegraphics[width=0.5\textwidth]{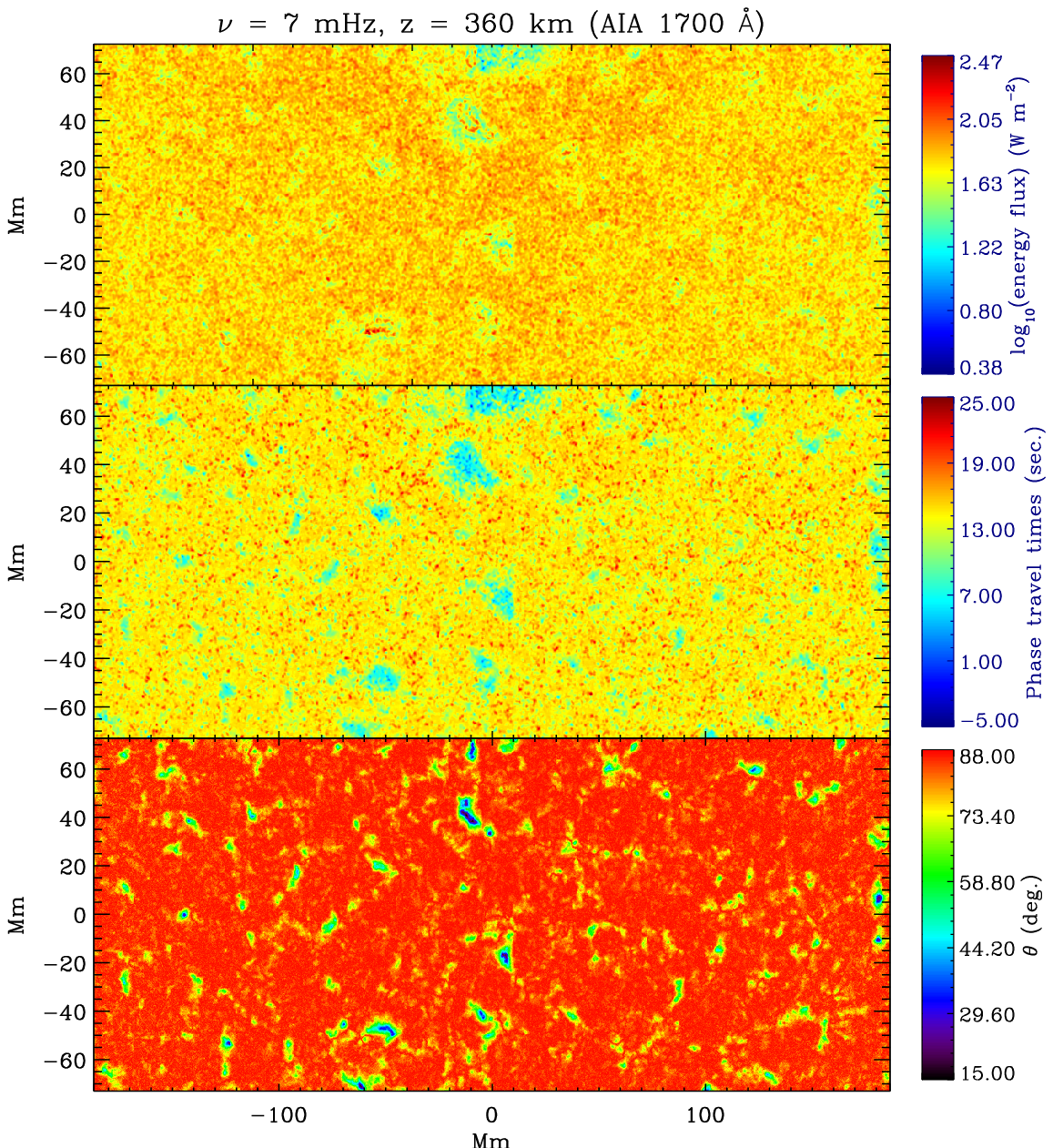}\\
\end{array}$
\caption{Maps of magnetic inclination angle $\theta$ (bottom panel), phase travel time $t_{\rm{ph}}$ (middle panel),
        and logarithm of energy flux $F$ over the quiet-network area of August 3, 2010 region shown in Fig.~\ref{fig1}. The
        $t_{\rm{ph}}$ and $F$ are for acoustic waves of frequency 3 mHz (left set of panels) and 7 mHz (right set of panels)
        propagating from 170 km (HMI line-core intensity $I_{\rm{co}}$) to 360 km (AIA 1700~{\AA} intensity $I_{\rm{uv1}}$).}\label{fig3}
\end{figure*}

The phase travel times, $t_{\rm{ph}}$, as a function of frequency $\nu$ are simply the time lags corresponding to $\delta\phi$
determined as,
\begin{equation}
\label{eq:4}
t_{\rm{ph}}(\nu) = \frac{\delta\phi(\nu)}{2\pi\nu}
\end{equation}
In our convention in Eq.(\ref{eq:1}), a positive $t_{\rm{ph}}$ (or $\delta\phi$) means that $I_{\rm{1}}$ leads $I_{\rm{2}}$ in phase
and hence corresponds to waves propagating upward and vice versa.
The photospheric continuum intensity $I_{\rm{c}}$ is dominated by granulation, and since
granules have lifetimes of the order of 5 to 10 minutes, the same as wave periods, they swamp the phase signals due to waves if we use it
in $I - I$ cross-spectra. Hence, we restrict to using the intensity pairs $I_{\rm{co}} - I_{\rm{uv1}}$ and $I_{\rm{co}} - I_{\rm{uv2}}$
to calculate cross-spectra and estimate $t_{\rm{ph}}$; these, then, correspond to following waves from upper photosphere
to chromospheric layers in the height range of 170 - 360 km and 170 - 430 km, respectively.
Between the photosphere and chromosphere, the $\delta\phi$ or $t_{\rm{ph}}$ receive a dominant negative contribution from
atmospheric gravity waves, which occupy the region of smaller frequencies bounded by the $f$-mode (surface-gravity mode) ridge
in the $k_{\rm{h}} - \nu$ dispersion diagram \citep[see, e.g.][and references therein]{2008ApJ...681L.125S}. These gravity waves
have downward phase propagation (hence negative $t_{\rm{ph}}$) while transporting energy upward. Since we are interested
in acoustic wave propagation, we first apply a helioseismic filter [a 3D Fourier filter in ($\nu,k_{\rm{x}},k_{\rm{y}}$)]
to the basic data cubes to remove the surface- and atmospheric-gravity waves \citep{2000PhDT.........9G,2004SoPh..220..381R}
before using them in the above cross-spectral analysis. 

\subsection{Estimating Energy Fluxes}\label{e_est}
\begin{figure*}[ht] 
\centering
\includegraphics[width=.5\linewidth]{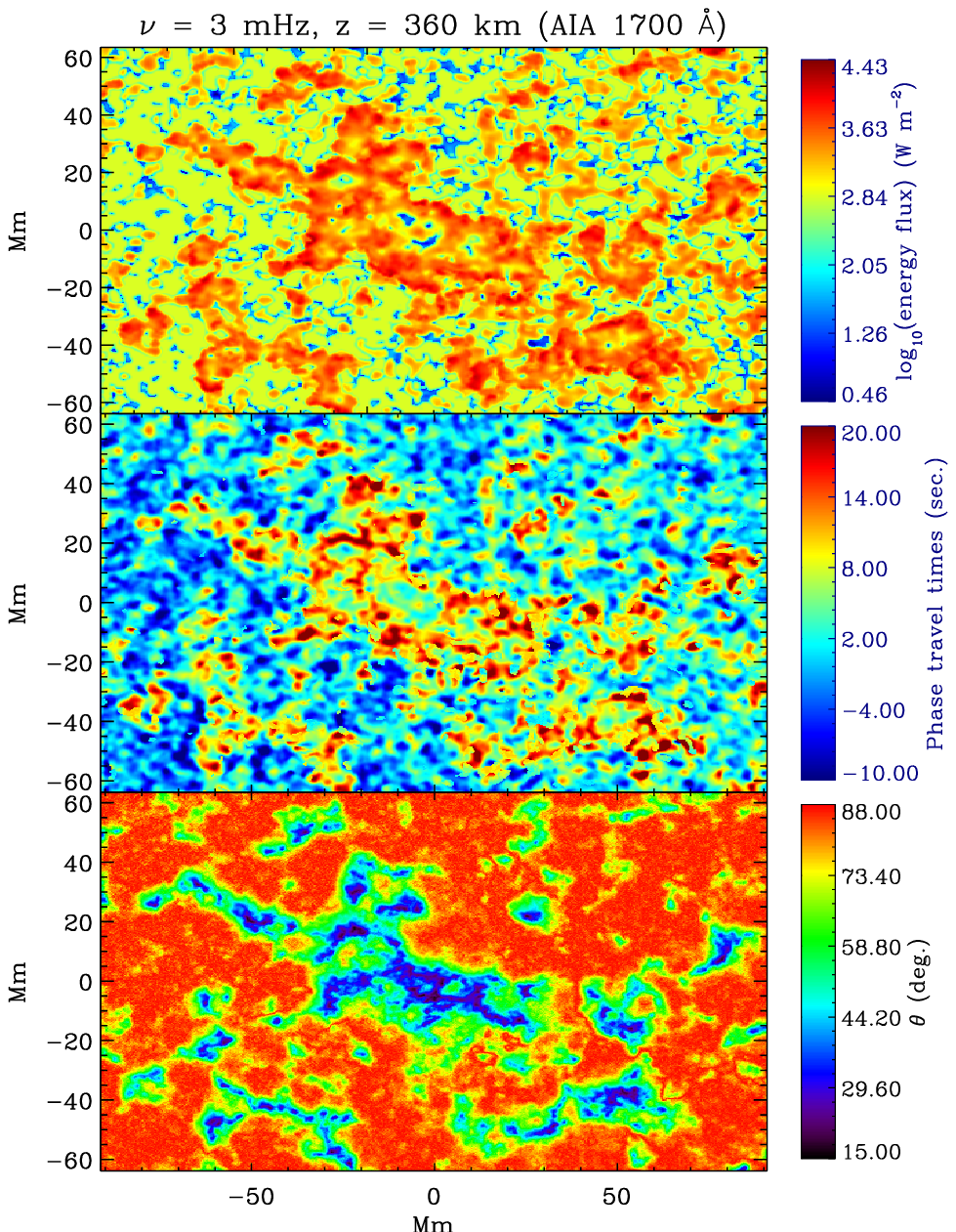}\includegraphics[width=.5\linewidth]{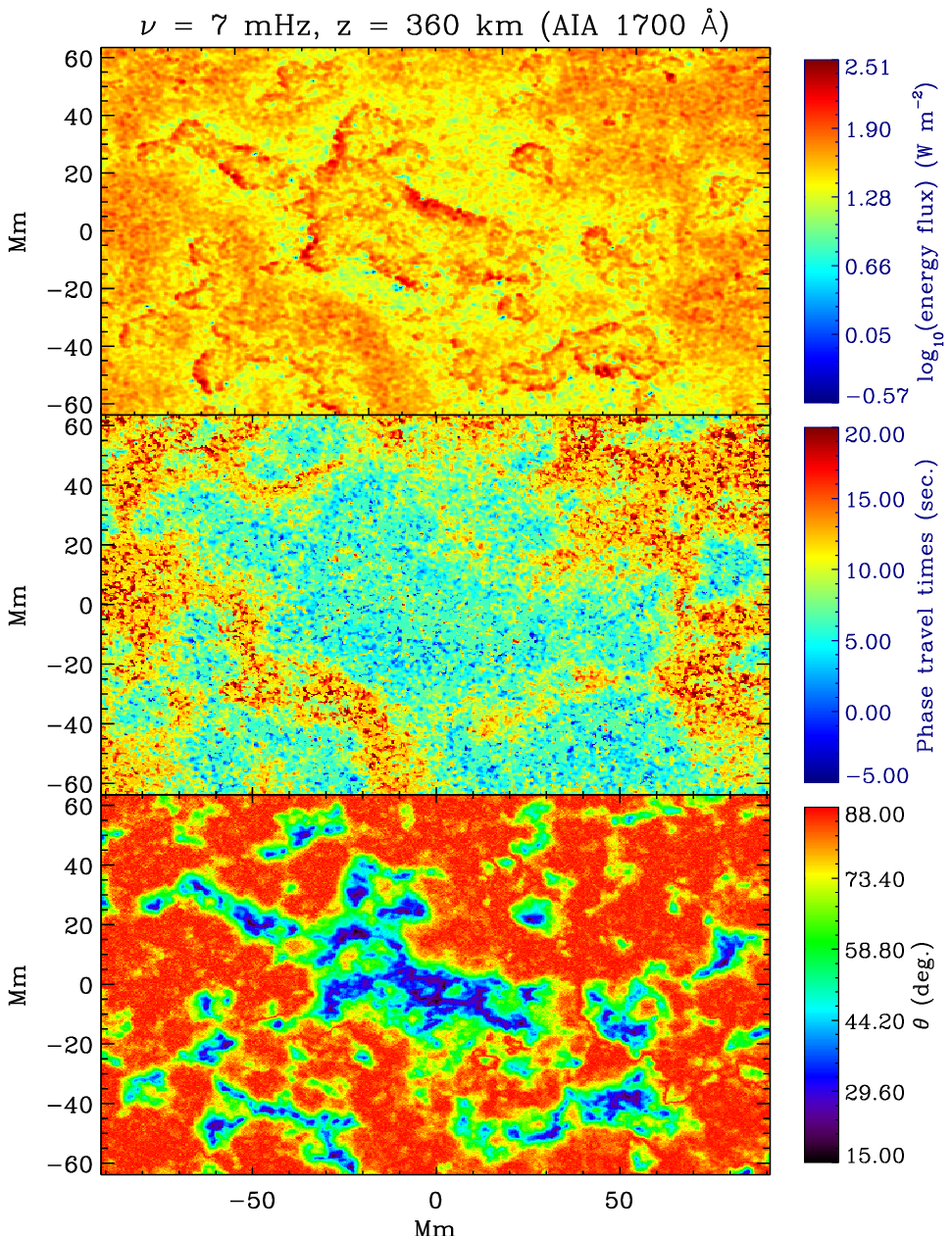}
        \caption{The same as Fig.~\ref{fig3} but for the plage region of August 3, 2010 shown in Fig.~\ref{fig1}.}\label{fig4}
\end{figure*}
Energy fluxes at heights 360 and 430 km corresponding to $I_{\rm{uv1}}$ and $I_{\rm{uv2}}$ are calculated
using the estimates of $t_{\rm{ph}}$ from the cross-spectra of intensity pairs $I_{\rm{co}} - I_{\rm{uv1}}$ and $I_{\rm{co}} - I_{\rm{uv2}}$
as follows (following the same basic principles as in \citet{2006ApJ...648L.151J}).
The basic expression that we use to determine energy flux $F$ at height $z$ is \citep{1973ApJ...184L.131C,2007SoPh..246...53D,2006ApJ...648L.151J},
\begin{equation}
\label{eq:5}
F(z)= P(z)\ v_{\rm{g}}(z)\ \rho(z),
\end{equation}
where $P$ is wave-velocity power, $\rho$ is the plasma density and $v_{\rm{g}}$ is the group velocity of the waves.
We use density values from the FAL 93 models \citep{1993ApJ...406..319F} the same way as \citet{2006ApJ...648L.151J}.
Group velocities at a given height are estimated from $t_{\rm{ph}}$ at that height using,
\begin{equation}
\label{eq:6}
v_{\rm{g}}(z) = \frac{c^{2}\ t_{\rm{ph}}(z)}{\triangle z},
\end{equation}
where $c$ is the sound speed, $\triangle z $ is the height difference that $t_{\rm{ph}}$ corresponds to. For $t_{\rm{ph}}$ from
the cross-spectral pairs $I_{\rm{co}} - I_{\rm{uv1}}$ and $I_{\rm{co}} - I_{\rm{uv2}}$, the $\triangle z $'s are 190 km and 260 km, respectively.
We use $c =$ 7 km/sec, which is appropriate for lower solar atmosphere. Note that the power $P$ in Eq.(\ref{eq:5}) has to be 
estimated at the height of $t_{\rm{ph}}$, but we have only the photospheric
velocities $v_{\rm{d}}$ and hence $P_{\rm{HMI}} = <v_{\rm{d}}^{2}>$. However, spatial map of $t_{\rm{ph}}$ allows us to estimate $P(z)$ 
using the height evolution of velocities for propagating and evansecent waves \citep[see, {\em{e.g.}}][]{JC-D03lecturenotes}:
conservation of mechanical energy flux dictates that the velocity amplitudes of propagating waves change as
$v(z) = v_{\rm{d}} \exp{[\triangle z_{\rm{v}}/2H_{\rho}]}$ while those of evanescent waves ($\nu < \nu_{\rm{ac}}$) decay as
$v(z) = v_{\rm{d}} \exp{[-\triangle z_{\rm{v}}(1-\nu^{2}/\nu_{\rm{ac}}^{2})^{1/2}/2H_{\rho}]}$, where $\triangle z_{\rm{v}} = z - z_{\rm{d}}$
is the height referenced to that of $v_{\rm{d}}$. Here, with $z_{\rm{d}} = $ 150 km for HMI velocity and 
$z=$ 360 km or 430 km for $I_{\rm{uv1}}$ or $I_{\rm{uv2}}$, we have $\triangle z_{\rm{v}} =$ 210 km or 280 km, respectively. 
To apply the former increasing amplitude against height, locations of upward propagation are identified as those having 
$t_{\rm{ph}} \geq \sigma^{\rm{t}}_{\rm{q}}$, the standard deviation of $t_{\rm{ph}}$ over all quiet (non-magnetic) locations with
HMI (time-averaged) $B_{\rm{tot}}$ less than 40 G. Since the propagation in height is mainly within magnetic fields, we need to consider 
the modification in height evloution caused by magnetic fields: solutions for longitudinal waves within vertically oriented
flux tubes give a height dependence $\exp{[\triangle z_{\rm{v}}/4H_{\rho}]}$ \citep{1981A&A....98..155S}, which arises from the
height variation of magnetic field. 
For power, just the squares of the above exponential height-evolution factors multiply $P_{\rm{HMI}}$ to give $P(z)$.
In addition, since we are largely tracking acoustic waves travelling along the magnetic field,
we correct for the underestimation of $v_{\rm{d}}$ (which is $LOS$ velocity) by $cos(\theta)$ factor due to magnetic field inclination $\theta$.
Estimates of $t_{\rm{ph}}$, and $P(z)$ from $P_{\rm{HMI}}$ as described above, at two heights 360 and 430 km corresponding 
to $I_{\rm{uv1}}$ and $I_{\rm{uv2}}$ are done in Fourier space at each frequency,
and hence the above procedure gives maps of energy fluxes, $F(x,y,\nu)$, at these two heights.
\begin{figure*}[ht] 
\centering
\includegraphics[width=0.5\linewidth]{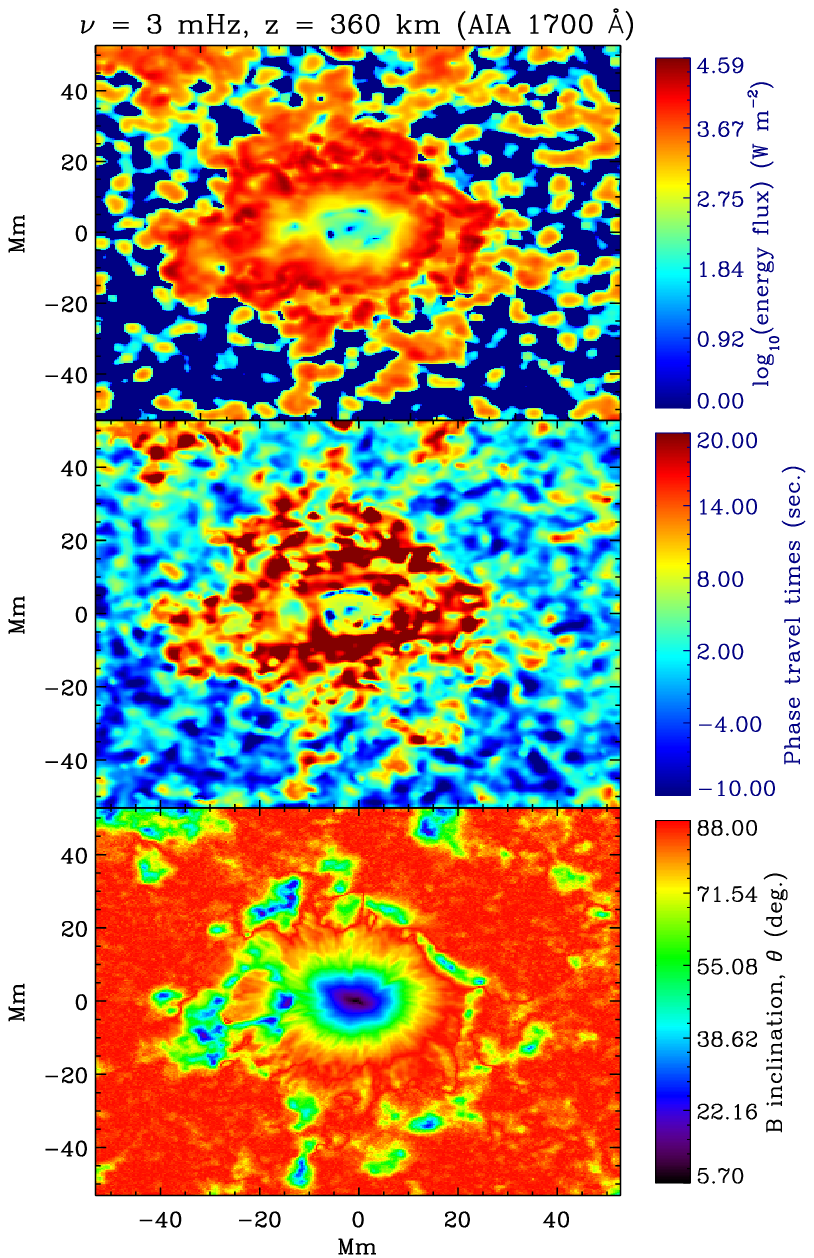}\includegraphics[width=0.5\linewidth]{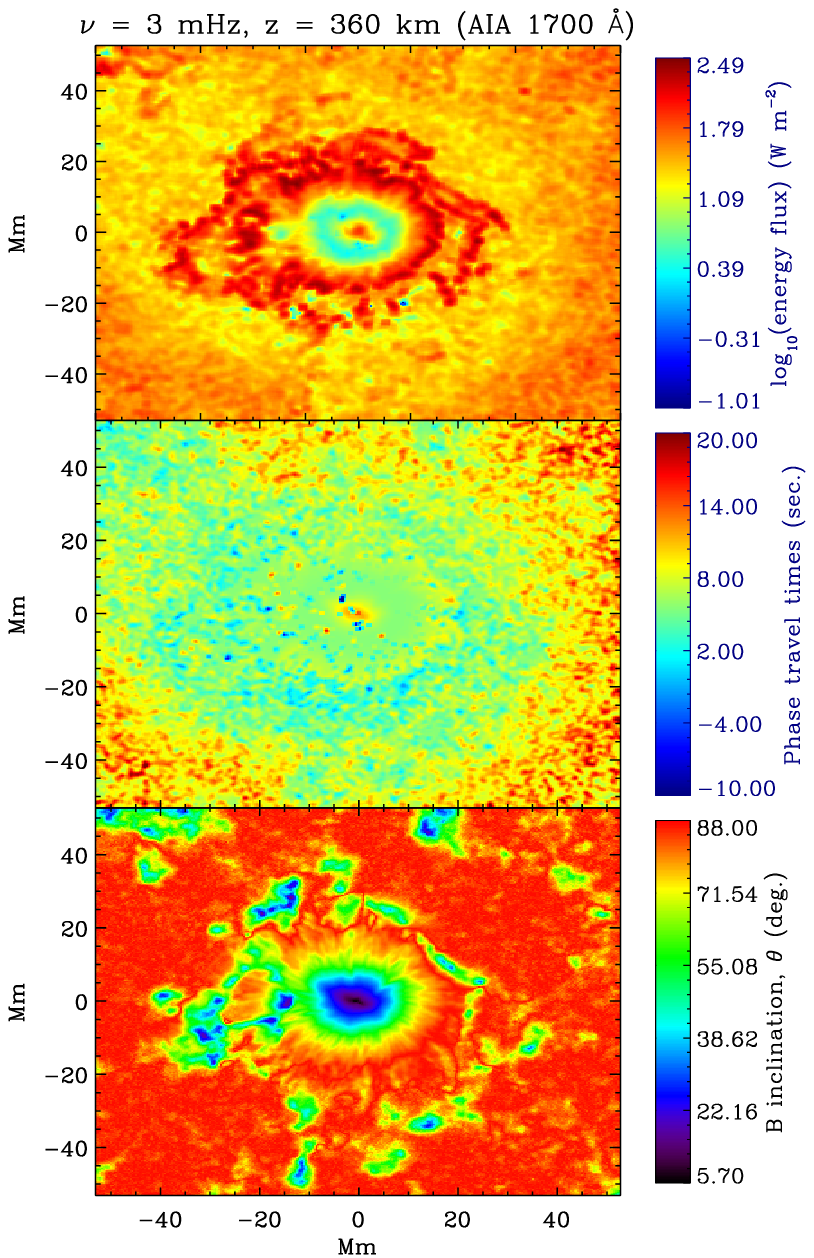}
        \caption{The same as Fig.~\ref{fig3} but for the spot region of August 3, 2010 shown in Fig.~\ref{fig1}.}\label{fig5}
\end{figure*}

Energy fluxes estimated by the above method, clearly, have a uncertainty range determined by that in $\triangle z$, $\triangle z_{\rm{v}}$
and that for $H_{\rho}$. While $\triangle z$ enters through $v_{\rm{g}}$, the other ones, through the exponential factors discussed above, 
determine $P(z)$. Moreover, spatial resolution of the data too play a role in energy estimates 
\citep{2006ApJ...648L.151J,2007ApJ...671.2154K,2009A&A...508..941B,2010A&A...522A..31B}. Much of the differences due to spatial resolution 
in the estimates of above authors have been for waves of frequencies larger than the quiet-Sun acoustic 
cut-off $\nu_{\rm{ac}}$ = 5.2 mHz. To assess the relative contribution of lower frequency, $\nu < \nu_{\rm{ac}}$, waves transmitted out 
by the quiet-Sun inclined magnetic fields, \citet{2006ApJ...648L.151J} calibrated their estimates by scaling them to match their 
higher-frequency (5.2 - 10 mHz) energy fluxes with that of \citet{2005Natur.435..919F}, who had higher spatial resolution 
(0.6~\arcsec resolution TRACE 1700~{\AA} and 1600~{\AA} observations) and had validated their estimates with numerical simulations. 
Over 5.2 - 10 mHz, quiet-Sun mean energy flux determined by these authors was 286.9 W m$^{-2}$ \citep{2006ApJ...648L.151J}. 
With the same spatial resolution and for the same chromospheric height probed by 1700~{\AA} (but here from SDO/AIA), 
our calculations yield very closely the same value (286.2 W m$^{-2}$, see Figure 11, left panel) when we use 
$\exp{[\triangle z_{\rm{v}}/4H_{\rho}]}$, with $H_{\rho} =$ 110 km, to determine wave velocities at height $\triangle z_{\rm{v}}$ 
above $z_{\rm{d}}$ using HMI velocities $v_{\rm{d}}$. 
The adopted value for $H_{\rho}$ agrees well with the FAL 93 models \citep{1993ApJ...406..319F}, which give $H_{\rho} \approx$ 109 km 
for the non-magnetic model (Model C) and $\approx$ 120 km for the magnetic model (Model P) for heights in the range of 150 - 450 km. 

It should be noted that the magnetic modification that halves the rate at which velocities increase over height, 
with 4$H_{\rho}$ scaling of height instead of 2$H_{\rho}$ in $\exp{[\triangle z_{\rm{v}}/4H_{\rho}]}$, is appropriate only 
for $\nu < \nu_{\rm{ac}}$ as otherwise all higher $\nu$ waves propagate in the non-magnetic atmosphere.
Hence, noting that power $P$ involves squares of the above exponential factors, a more realistic value for the 5.2 - 10 mHz 
mean energy flux in our calculations would be $\exp{[\triangle z_{\rm{v}}/2H_{\rho}]} \times $286.2 = 743.0 W m$^{-2}$, for 
$\triangle z_{\rm{v}} = $210 km ({\it i.e.}, z = 360 km, AIA 1700~{\AA}). On the other hand, as discussed in \S\ref{phcoh}, 
we are in the regime of $\tau_{\rm{R}} < \tau_{\rm{S}}$, and hence RD will amount to a reduction in amplitudes over height
faster than $\exp{[\triangle z_{\rm{v}}/2H_{\rho}]}$, and this possibly explains the lower value of 286.9 W m$^{-2}$ obtained
by \citet{2005Natur.435..919F}.
We stress, however, the importance of accounting for the magnetically confined propagation of acoustic waves for
$\nu < \nu_{\rm{ac}}$ through the use of $\exp{[\triangle z_{\rm{v}}/4H_{\rho}]}$ in calculating velocities
over height $\triangle z_{\rm{v}}$. The resulting mean energy fluxes over 2 - 5.2 mHz range are presented and discussed in \S\ref{res3}. 
We further note that, since extending the above to quiet-Sun propagation at $\nu > \nu_{\rm{ac}}$ leads to an underestimation 
by nearly a factor of $e$, but matches with that of \citet{2005Natur.435..919F} without any additional scaling as done by 
\citet{2006ApJ...648L.151J} to calibrate their estimates, our estimates for the lower frequencies are likely to be conservative
or lower limits.

\section{Results and Discussions}\label{main} 
\subsection{Maps of Phase Travel-times and Energy Fluxes}\label{res1}

The phase travel-times $t_{\rm{ph}}$ and coherence $C$, as described in \S \ref{phcoh}, are estimated at the two heights, 360 and 430 km,
over the two regions shown in Figure 1. Before analysing the variations of $t_{\rm{ph}}$ against $B_{tot}$ and $\theta$, we first
check if the quiet-Sun changes in $t_{\rm{ph}}$ against frequency $\nu$ confirm with the expected behaviour without 
much contamination due to height-dependent RD, as discussed in \S\ref{phcoh}. Within the
quiet-Sun areas marked in Figure 1 with blue boundaries, we separate the magnetic and non-magnetic pixels with the criterion
of $B_{\rm{tot}} < $40 G for the latter and then average $t_{\rm{ph}}$ and $C$ over these two groups of pixels. The results are
shown in Figure 2. The $\nu$ variation of quiet-Sun travel times (or phase difference against height) shows the familiar
trend, with the main p-mode band showing the evanescent behaviour: smaller or close-to-zero $t_{\rm{ph}}$ within the 2 - 5 mHz 
with the smallest values at 3 mHz, and correspondingly large coherence (peak values of ~0.8 near 3 mHz). The magnitudes and $\nu$ dependence
of quiet-Sun $t_{\rm{ph}}$ obtained here match very closely the recent estimates, for the same height range, from multi-height 
velocity data \citep{2016ApJ...819L..23W}. 
Overplotted in Figure 2 are the $\nu$ variations of $t_{\rm{ph}}$ and $C$ averaged over magnetic pixels. The magnetic field assisted
propagation is clear in the enhanced low-frequency travel times -- this is the turn-on of propagation due to reduction of
acoustic cut-off frequency by the inclined magnetic fields \citep{2006ApJ...648L.151J}; we explore in detail this behaviour
in the rest of the paper, including revised estimates of energy fluxes.
Note that, in Figure 2,
we have not separated the phase shifts over inclined and near-vertical fields as in \citet{2006ApJ...648L.151J} but
do a detailed analysis of dependence on $\theta$ in \S\ref{res2} (see Figure 6), where a good agreement with \citet{2006ApJ...648L.151J}
is found for inclined magnetic fields while uncovering some new features in the nearly vertical small-scale fields. 
As argued in \S\ref{phcoh}, the above described agreement between $I(z_{1}) - I(z_{2})$ phase shifts derived here 
and the $V(z_{1}) - V(z_{2})$ phase shifts of \citet{2016ApJ...819L..23W} and \citet{2006ApJ...648L.151J} for the quiet-Sun evanescent 
and the magnetic inclination-assisted propagation, respectively, in the frequency range 2 - 5 mHz shows that the height dependence of RD plays negligible roles.
We stress that only the height dependence of RD and hence contamination or artificial phase shift due to it is negligible, but RD itself is 
significant as $\tau_{\rm{R}} < \tau_{\rm{S}}$ in the height ranges considered and hence the real physical effect of lowered acoustic cut-off
is still possible. This we believe is why the quiet-Sun curve (Figure 2), similar to the results of \citet{2016ApJ...819L..23W}, 
exhibits significant propagation (non-zero $t_{\rm{ph}}$) down to about $\approx$ 4 mHz, while the usual theoretical value for acoustic cut-off is
$\nu_{\rm{ac}}$ = 5.2 mHz.
It is to be noted that the coherence values, on average, 
remain largely above 0.5 over quiet and magnetic pixels (except at high frequencies), and hence the estimates of $t_{\rm{ph}}$
can reliably be attributed to waves; for estimates of mean energy fluxes (over observed areas) we apply a correction for
pixels with low coherence (see next Section).

We show in Figures 3 - 5 the 3 and 7 mHz maps of travel-times $t_{\rm{ph}}$ and energy fluxes $F$, along with that of 
magnetic inclination $\theta$, for the quiet-Sun, plage and spot areas of the {\em 03 Aug 2010} region (marked in Figure 1).
These maps portray striking variations and patterns as a function of frequency and background magnetic structures.
At frequencies lower than $\nu_{\rm{ac}}$ (= 5.2 mHz), it is readily recognised that $t_{\rm{ph}}$ and $F$ are concentrated within the 
the discrete magnetic field elements of the quiet-Sun ({\it left panel} of Figure 3), confirming what
\citet{2006ApJ...648L.151J} described as magneto-acoustic portals. In the plage ({\it left panel}, Figure 4) and
sunspot ({\it left panel}, Figure 5) regions, however, the lower frequency enhancements are larger in the more inclined boundary
regions of magnetic flux concentrations. At frequencies larger than $\nu_{\rm{ac}}$ (refer to the 7 mHz maps in the {\it right panels} 
of Figures 3 - 5), all magnetic flux concentrations exhibit a nulling or much reduced $t_{\rm{ph}}$ around them, {\it i.e.} in
the inclined canopy regions, and correspondingly there is much reduced upward energy flux. This phenomenon around small-scale
magnetic fields at chromospheric heights was observed in high-resolution by \citet{2007A&A...461L...1V}. At the photospheric and
upper photospheric levels, these signatures in $t_{\rm{ph}}$ spatially coincide with the so called {\it acoustic halos}
that represent enhanced oscillation power \citep[see,][and references therein]{2013SoPh..287..107R}. This reduction or nulling
of $t_{\rm{ph}}$ has been modeled and interpreted as due to mode-conversion assisted horizontal \citep{2012A&A...538A..79N} or
downward propagation of fast magnetoc-acoustic waves at the canopy \citep{2016ApJ...817...45R}. Another feature in the larger
flux concentrations, especially in the sunspot, is the pattern of increasing frequencies from the outer edge (3 mHz in more inclined fields)
towards the central ($>$ 5 mHz in the nearly vertical umbral fields) region (refer to the {\it right panels} in Figure 5). This feature
has also been observed earlier in different contexts \citep{2016ApJ...830L..17Z,2016NatPh..12..179J} -- a more
detailed examination of this and its relation to $\theta$ and $\nu^{B}_{\rm{ac}}$ is presented in the next Section. 
\begin{figure*}[ht] 
$\begin{array}{rl}
    \includegraphics[width=0.5\textwidth]{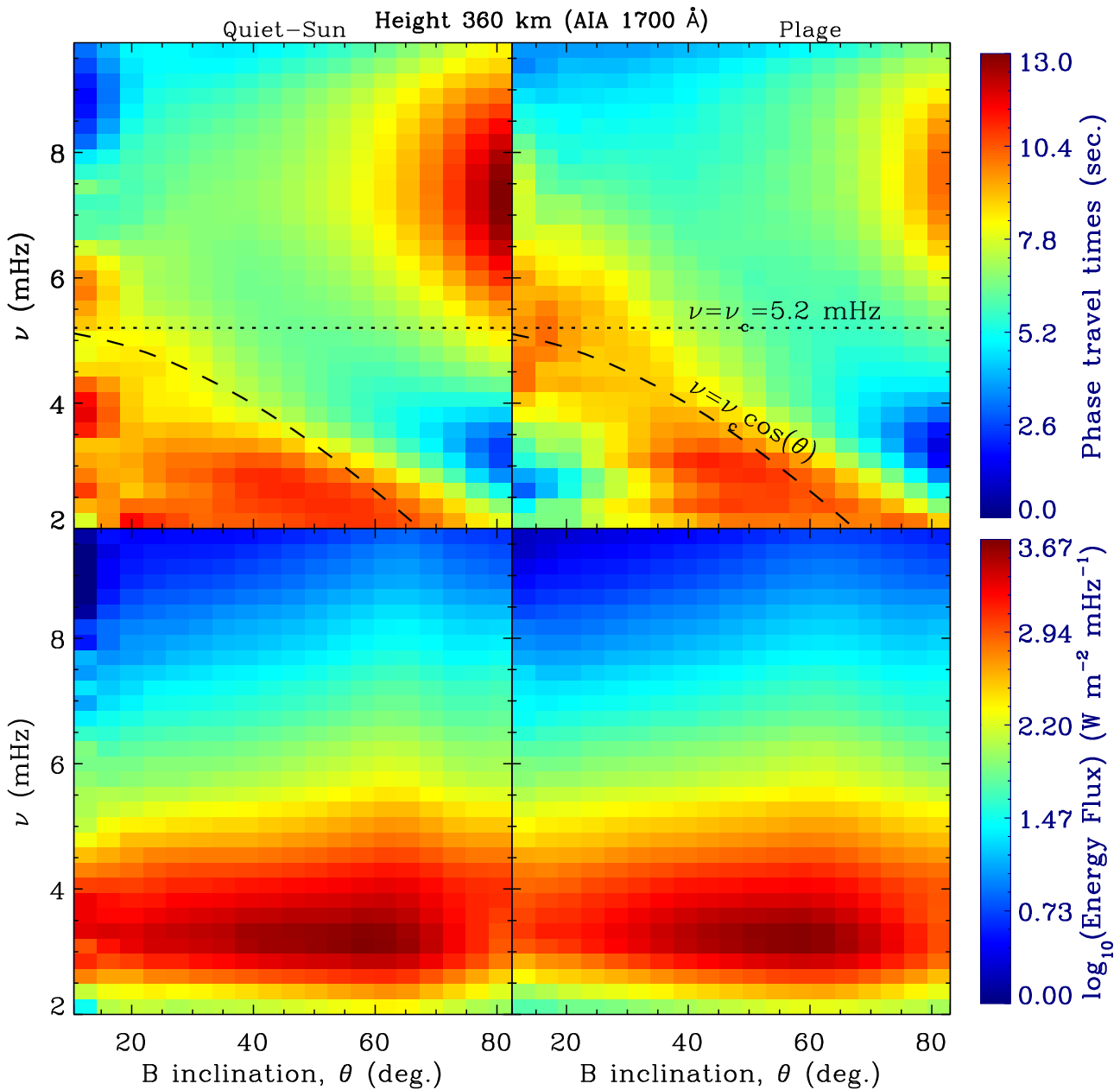} &
    \includegraphics[width=0.5\textwidth]{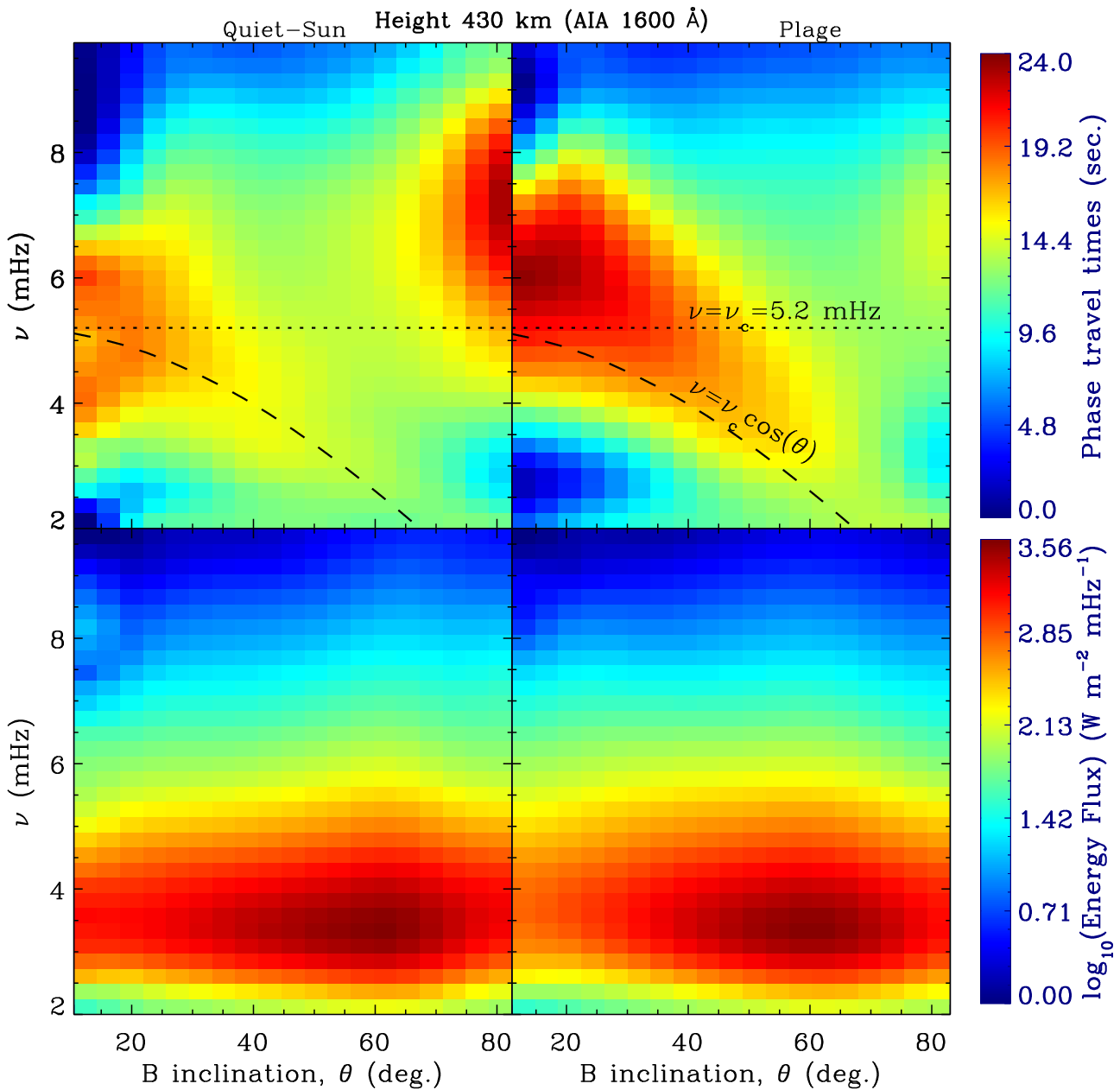}\\
\end{array}$
\caption{Phase travel times, $t_{\rm{ph}}$, and logarithm of energy fluxes, $F$, against magnetic inclination angle, $\theta$,
        and wave frequency, $\nu$, compared for quiet and plage regions; left set of panels is for AIA 1700~{\AA} height (360 km) and
        that on the right is for AIA 1600~{\AA} height (430 km).}\label{fig6}
\end{figure*}

It is to be noted that, although we have filtered out the low-frequency atmospheric gravity modes ({\it cf.,} \S\ref{phcoh}),
the maps of $t_{\rm{ph}}$ (middle panels of Figures 3 - 5) still show significant areas of negative values (corresponding
to downward propagation) at low frequencies (2 - 4 mHz). In general, $k - \nu$ cross-spectra of two-height oscillation data
(in intensities as well as velocities) have been known to show negative phase-differences between evanescent 
({\it i.e.}, ~ zero phase-difference) p mode ridges, but their origin is not understood \citep[see,][and references therein]
{2001A&A...379.1052K,2013SoPh..284..297S,2014SoPh..289.3457N}. The spatial average of $t_{\rm{ph}}$ over all quiet non-magnetic pixels, however,
show the nearly evanescent behaviour in 2 - 4 mHz range (see Figure 2). In the top panels showing energy fluxes $F$ in Figures 3 and 4, 
for the purpose of display, we have replaced all the negative $t_{\rm{ph}}$ locations within the non-magnetic quiet area 
by that $F$ estimated using the quiet-Sun average values shown in Figure 2. 
\subsection{Magnetic Inclination and Strength Dependences of Wave Propagation}\label{res2}

Effects of magnetic fields on wave-propagation and energy fluxes are studied
by deriving the variation of $t_{\rm{ph}}$ and $F$ against $\theta$ and $B_{tot}$ at each $\nu$. For this purpose, 
we group (and average) pixels in the 2D maps $t_{\rm{ph}}(x,y)$ and $F(x,y)$ (Figures 3 - 5) over 
bins of 4{\textdegree} in $\theta$ and 30~G in $B_{\rm{tot}}$. Although, on average, coherence $C > 0.4$ over much of the frequency ranges
considered here, there are significant quiet- as well as magnetic locations with lower $C$, and hence we select
only those locations where $C > 0.4$ for deriving the dependences on $\theta$ and $B_{tot}$.
The results are 2D functions $t_{\rm{ph}}(\theta,\nu)$, $t_{\rm{ph}}(B_{\rm{tot}},\nu)$, $F(\theta,\nu)$, and $F(B_{\rm{tot}},\nu)$
shown in Figures 6 - 8: quiet and plage regions' results are plotted in the same set of panels in Figures 6 and 7, the spot region is 
plotted separately in Figure 8.
\begin{figure*}[ht] 
$\begin{array}{rl}
    \includegraphics[width=0.5\textwidth]{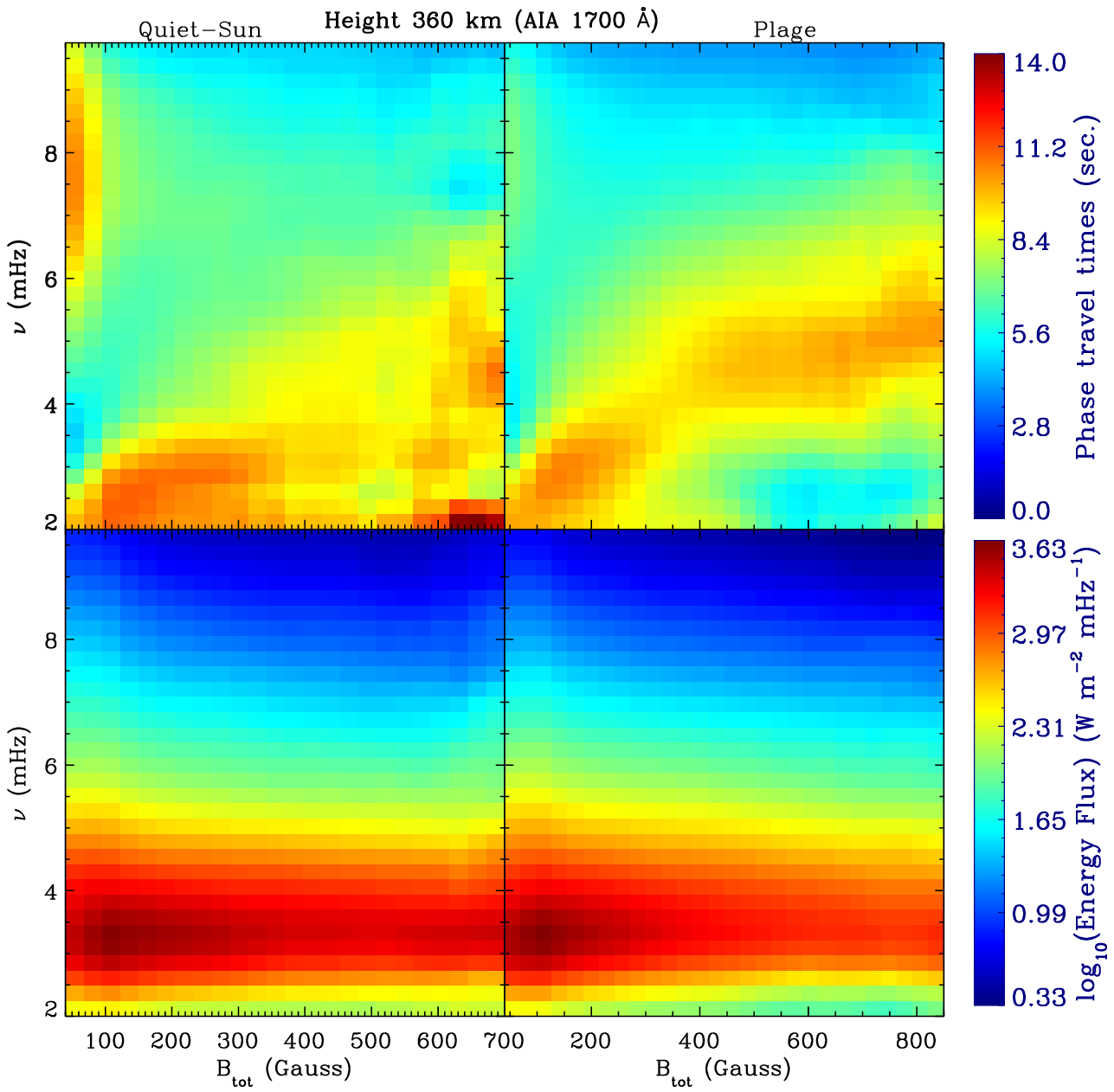} &
    \includegraphics[width=0.5\textwidth]{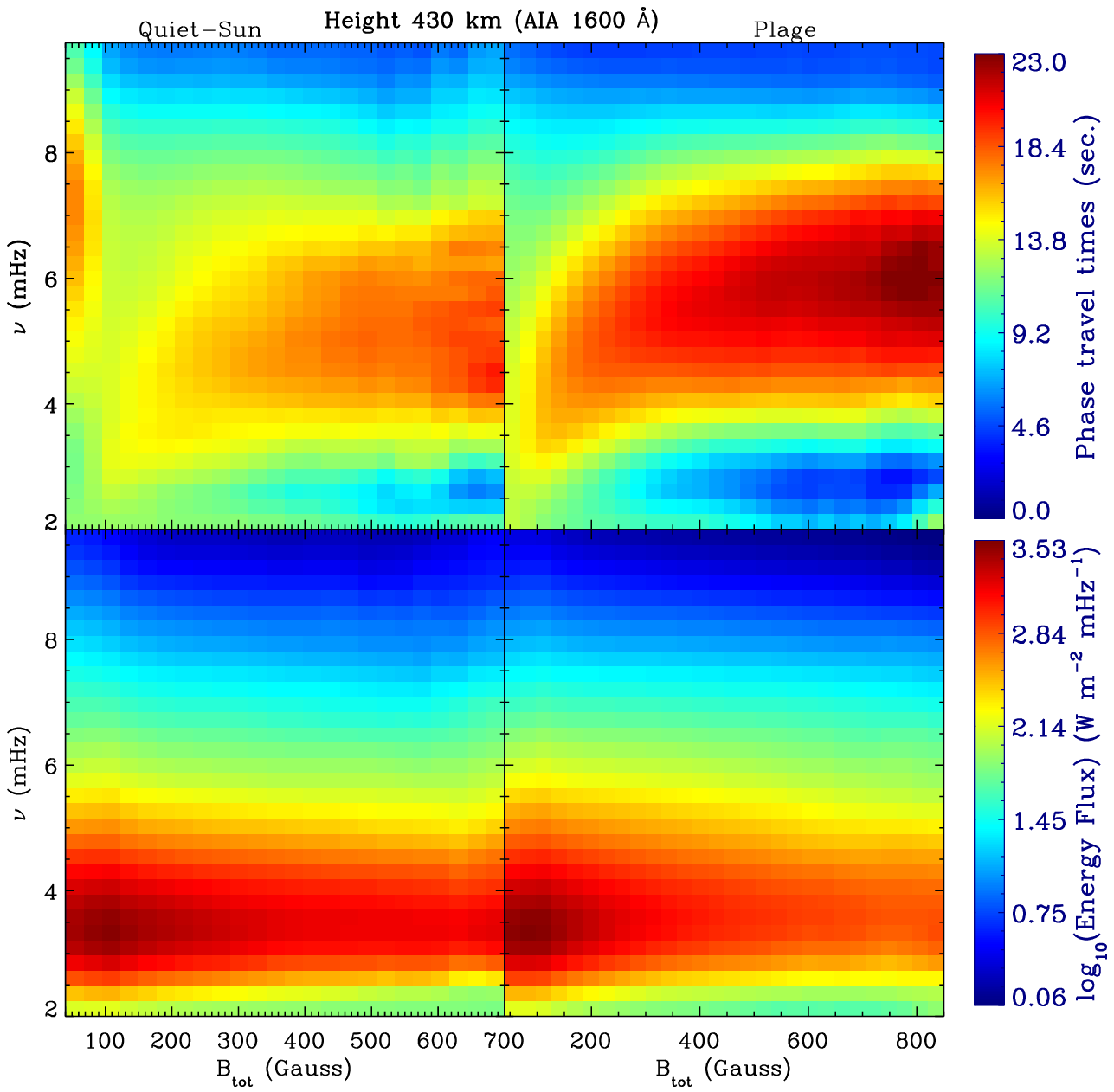}\\
\end{array}$
\caption{Phase travel times, $t_{\rm{ph}}$, and logarithm of energy fluxes, $F$, against total magnetic field strength, $B_{tot}$, and
        wave frequency, $\nu$, compared for quiet and plage regions; left set of panels is for AIA 1700~{\AA} height (360 km) and
        that on the right is for AIA 1600~{\AA} height (430 km).} \label{fig7}
\end{figure*}
The two sets of panels in Figures 6 and 7 show $t_{\rm{ph}}$ and $F$ at heights 360 km (1700~{\AA}) and
430 km (1600~{\AA}); in each set the upper sub-panels show $t_{\rm{ph}}(\theta,\nu)$ or $t_{\rm{ph}}(B_{\rm{tot}},\nu)$, and the
lower ones $F(\theta,\nu)$ or $F(B_{\rm{tot}},\nu)$ for the quiet and plage regions. 

\subsubsection{Phase Travel-times}\label{res2.ph}
The inclination $\theta$ of magnetic fields to the direction of gravity, as referred to earlier in \S\ref{dat}, reduces
the effective gravity that the fluid tied to the magnetic field is subject to by a factor $\approx cos(\theta)$ and
hence the acoustic cut-off frequency, $\nu_{\rm{ac}}$, that is proportional to the acceleration due to gravity $g$ is similarly reduced \citep{1977A&A....55..239B}.
\begin{figure*}[ht] 
$\begin{array}{rl}
    \includegraphics[width=0.5\textwidth]{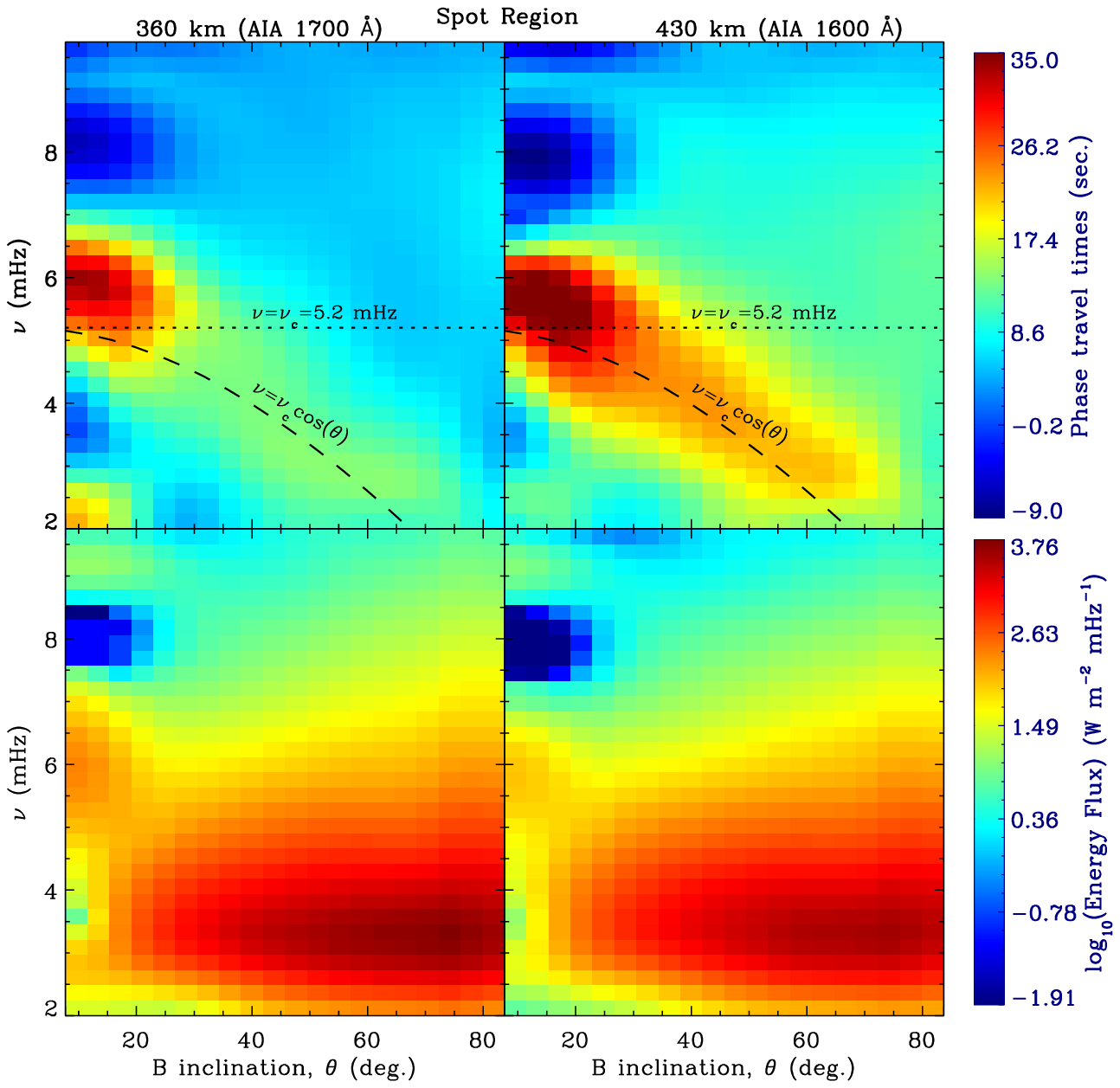} &
    \includegraphics[width=0.5\textwidth]{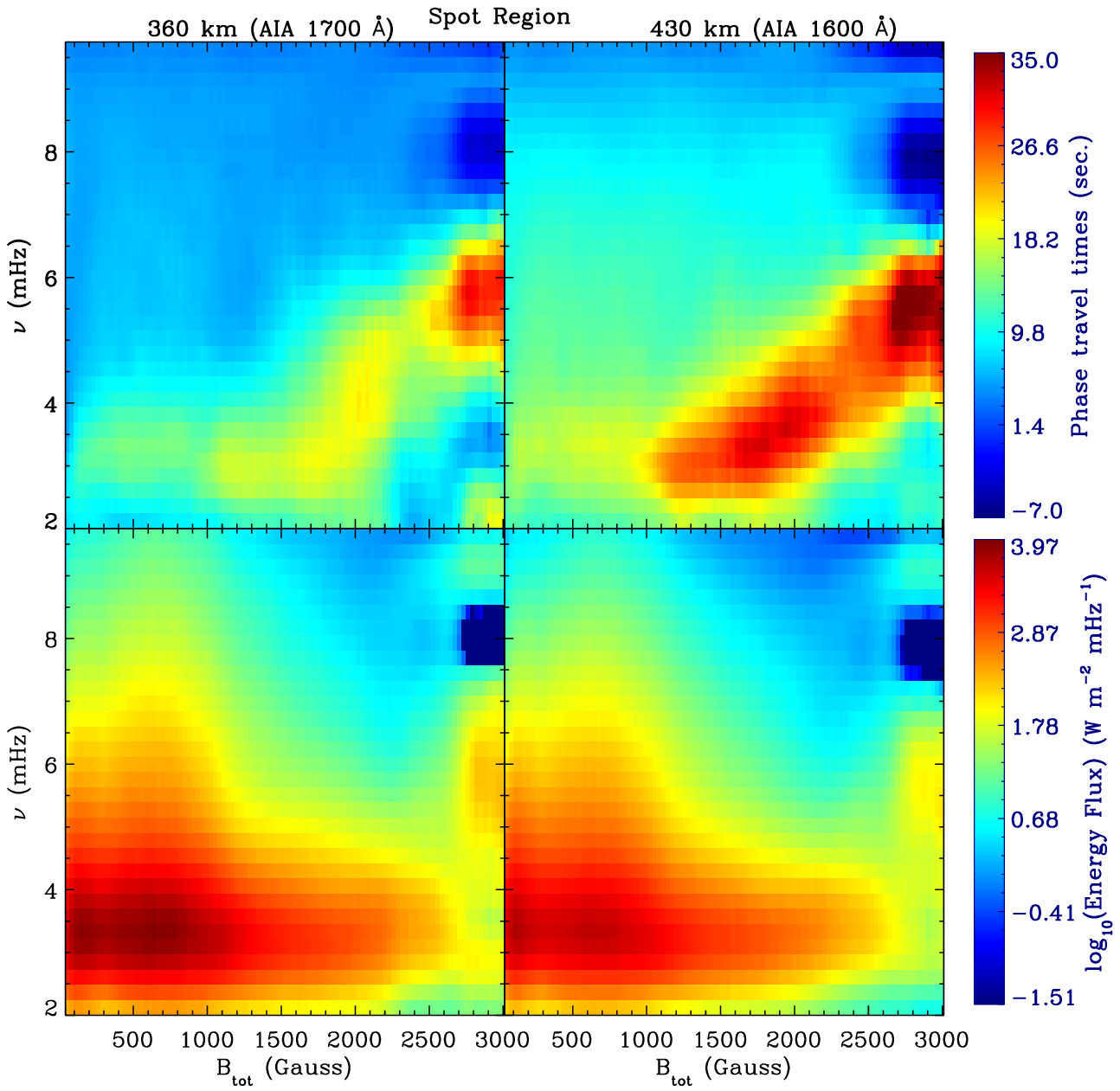}\\
\end{array}$
        \caption{Phase travel times, $t_{\rm{ph}}$, and logarithm of energy fluxes, $F$, against magnetic inclination angle $\theta$
        and wave frequency $\nu$ (left set of panels), and against total magnetic field strength $B_{tot}$ and wave frequency $\nu$ (right set of panels)
        for the spot region.} \label{fig8}
\end{figure*}
We have overplotted the theoretical minimum $\nu^{B}_{\rm{ac}}=\nu_{\rm{ac}}cos\theta$ due to magnetic feilds in 
the maps of $t_{\rm{ph}}(\theta,\nu)$ in Figures 6 - 8 as dashed curves. The region of smaller 
$\theta$ and $\nu$ bounded by this curve in the $\theta - \nu$ plane, in principle, is not accessible for wave-propagation
via reduction in $\nu_{\rm{ac}}$ due to magnetic inclination $\theta$. However, Figure 6 shows that,
strikingly, almost the whole of this region of low $\nu$ and $\theta < $ 60\textdegree~is filled with propagating waves in the case of 
lower height (360 km, 1700~{\AA}) $t_{\rm{ph}}(\theta,\nu)$ corresponding to quiet-network (the top left sub-panel in Figure 6;
note also the different ranges of color-scales for $t_{\rm{ph}}$ in the left and right set of panels).
It is to be noted that locations where $\theta > $80\textdegree~correspond to very weak magnetic fields or completely quiet,
and the $\theta - \nu$ behaviour of $t_{\rm{ph}}$ in this region (in Figures 6 and 7) is, overall, consistent with
that expected in quiet-Sun: propagation at frequencies $>$ 5 mHz, largely, and evanescent behaviour otherwise, with a caveat that,
similar to the results of \citet{2016ApJ...819L..23W}, the quiet-Sun acoustic cut-off is lower than the usual
theoretical value of $\nu_{\rm{ac}}$ = 5.2 mHz with non-zero propagation down to about $\approx$ 4 mHz at $\theta > $80\textdegree.

The above described propagation in the low $\nu$ (2 - 4 mHz) and largely in the less inclined (from nearly vertical
$\theta \approx $10\textdegree~to $\theta < $60\textdegree) region over the quiet magnetic-network is distinct from that could be explained by
any modification or reduction of acoustic cut-off. The corresponding signatures are also
clear in $t_{\rm{ph}}(B_{\rm{tot}},\nu)$ shown in the top left sub-panels of Figure 7: stronger ($>$ 500 G), less inclined fields
show enhanced $t_{\rm{ph}}$. Further, over height, this propagation region diminishes both in $t_{\rm{ph}}(\theta,\nu)$ and
$t_{\rm{ph}}(B_{\rm{tot}},\nu)$: at height 430 km (1600~{\AA}) the $t_{\rm{ph}}$ in 2 - 4 mHz region, 
for both the quiet-network and the plage, do not show an increase expected over height, and in fact
there is a slight reduction at the lowest frequencies. This is possible only if all of the propagating waves seen at the lower height 
do not reach the higher location, in turn implying wave steepening and dissipation \citep{2001A&A...379.1052K}. 
To see the changes over height clearly, we have plotted the height difference in 
$t_{\rm{ph}}$, {\it i.e.,} difference between the 430 km (1600~{\AA}) and 360 km (1700~{\AA}) $t_{\rm{ph}}$ shown in the top left panels
of Figure 6, averaged over $\theta < \theta_{\rm{ac}}$ = $cos^{-1}(\nu/\nu_{\rm{ac}})$
and over $\theta \ge \theta_{\rm{ac}}$ in the left panel of Figure 9: it is clear that waves in the $(\theta,\nu)$ space allowed
via reduced $\nu_{\rm{ac}}$ due to magnetic inclination propagate to heights above 430 km (dashed curve), whereas that in the
region of smaller $\theta$ and $\nu$ bounded by $\nu^{B}_{\rm{ac}}$ have smaller $t_{\rm{ph}}$ (solid curve) for the
same height difference indicating wave steepening. This then implies that waves propagating in the region 
$\theta < \theta_{\rm{ac}}$ evolve faster to shock at lower chromospheric heights.
A similar $\theta$-integration of energy fluxes $F$ (corresponding to the lower left panels of Figure 6) and ratios of them are 
shown in the right panel of Figure 9 and are discussed in \S\ref{res3}.
\begin{figure*}[ht] 
\centering
\includegraphics[width=0.7\textwidth]{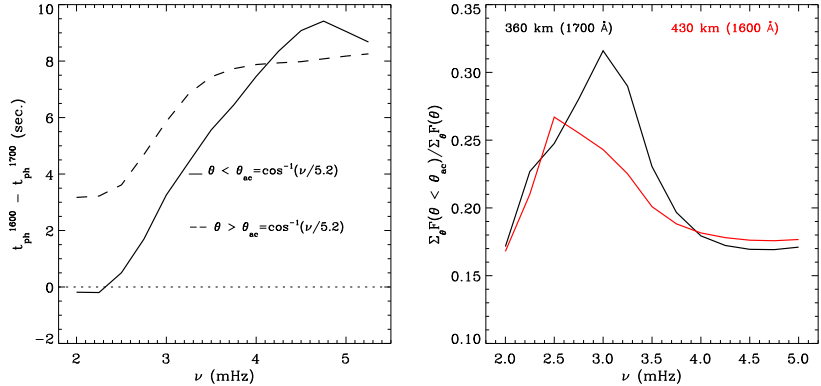}
\caption{{\it Left panel}: differences in $t_{\rm{ph}}$, averaged over $\theta < \theta_{\rm{ac}}$ = $cos^{-1}(\nu/\nu_{\rm{ac}})$ and
over $\theta \ge \theta_{\rm{ac}}$, of 430 km (1600~{\AA}) and 360 km (1700~{\AA}) heights shown in the top left panels
of Figure 6. {\it Right panel}: ratios of energy fluxes $F$ integrated over $\theta < \theta_{\rm{ac}}$ to that integrated over the whole range.
        See \S\ref{res3} for details.}\label{fig9}
\end{figure*}

Given that the small-scale magnetic flux elements are highly dynamic, one can argue that the propagation region of $\theta - \nu$
in Figure 6 (top left panel) is actually due to the intermittent excursions to large inclinations (beyond the dashed curve) of a flux element, while
its average $\theta$ (used in the analyses here) remains low. To check this, we plot the standard deviation of $\theta$ over
the full time-length of data used (at 720 sec. cadence) against the mean $\theta$ in Figure 10: the light blue band shows the extent
or distribution of standard deviation over the duration of observation at each location that has the mean $\theta$ (shown in the $x-axis$). The green
dots running at the center of the blue band are the RMS of standard deviation at 2\textdegree~interval. Up to about $\theta$ = 30\textdegree~the 
maximum standard deviation is less than 20\textdegree, and hence flux elements with average inclinations less than 30\textdegree~still do not
reach inclinations $\theta >$ 50\textdegree~to allow propagation at frequencies 3.3 mHz and lower. Hence, the dynamic nature of flux elements
is still not sufficient enough to cause the low $\nu$ (2 - 4 mHz) and $\theta < $40\textdegree~propagation seen in Figure 6. 
\begin{figure*}[ht] 
\centering
\includegraphics[width=0.7\textwidth]{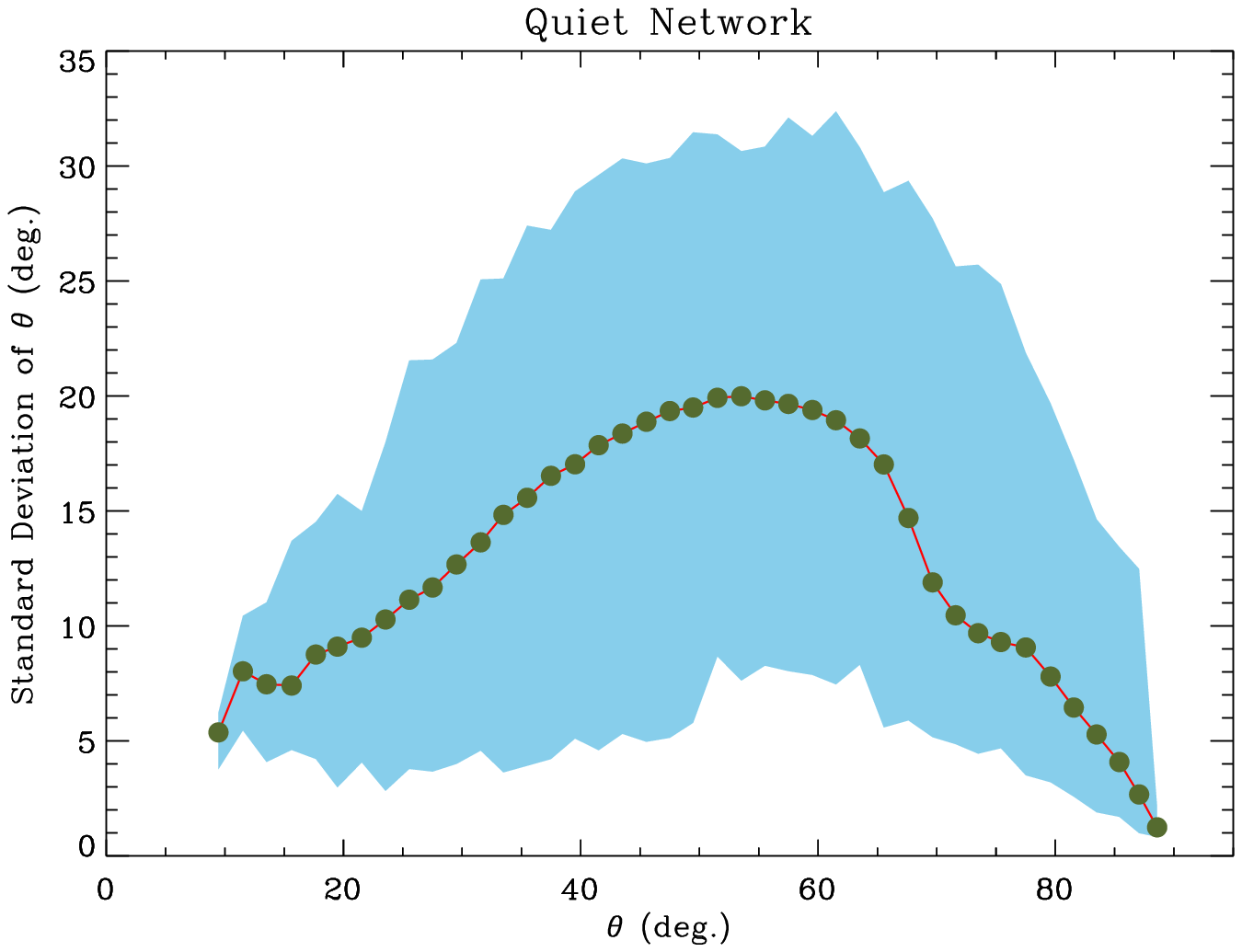}
\caption{The standard deviation of $\theta$ against the mean $\theta$ in the quiet magnetic-network. The light blue band
        shows the extent of standard deviation over the duration of observation at each location that has the mean $\theta$ (shown in the $x-axis$).}
        \label{fig10}
\end{figure*}

Although we have verified in \S\ref{res1} that RD has not contaminated ({\em i.e.,} introduced artifacts in) the quiet-Sun $t_{\rm{ph}}$
confirming our identification in \S\ref{phcoh} that $\tau_{\rm{R}}$ is nearly constant over the height ranges considered,
we still need to address if the nearly vertical network flux tubes present suitable physical conditions so different from the quiet-Sun
that either (i) artifacts of height dependent RD, or (ii) a large reduction in acoustic cut-off due to strong RD \citep{1978SoPh...59...49G,
1983SoPh...87...77R,2006ASPC..358..465C,2008ApJ...676L..85K} could be responsible for the above 2 - 4 mHz propogation. 

In case (i), we have the situation that the 2 - 4 mHz waves inside vertical tubes are in fact evanescent but the detection method picks up a 
positive phase-shift as an artifact mimicking upward propagation; this can happen only if RD increases via a decreasing $\tau_{\rm{R}}$ over 
height inside the flux tube, in contrast to the opposite trend of slightly increasing $\tau_{\rm{R}}$ in the quiet-Sun over the height ranges 
covered here \citep{1982ApJ...263..386M,2011MNRAS.417.1162N,2013SoPh..284..297S}. 
Further, as seen in Figure 6 (top left panel), the 2 - 4 mHz phase shifts over $\theta < $40\textdegree~is nearly two to five times that in the 
quiet-Sun and as we discuss below the vertical fields within the plage region do not show such propagation (Figure 6, top right panel). 
Hence the decrease in $\tau_{\rm{R}}$ over height within isolated flux tubes has to be of large magnitude and vastly different from not 
only that in quiet-Sun but also from plage region. This highly contrived situation within only the small-scale network elements is 
highly unlikely. 

For case (ii), although the requirement is less stringent in that it concerns only the physical effect depending on 
the magnitude of RD, a large reduction in $\tau_{\rm{R}}$ within network flux tubes as compared to quiet-Sun or even 
the vertical fields of plages is still needed.
We note that \citet{2006ASPC..358..465C} and \citet{2008ApJ...676L..85K} have reported observations with models based
on case (ii) but over heights between 400 - 1500 km; these heights mostly lie above the region covered in this work, and further since the
vertical flux tubes expand rapidly over height to fill the chromosphere at these heights, it becomes difficult to differentiate waves propagating 
due to RD with those due to magnetic inclination. Moreover, these authors report phase delays peaking around 3 mHz, which they achieve
as a reduced cut-off frequency within magnetic tubes due to RD in their models with $\tau_{\rm{R}}$ = 10 sec as compared to 50 sec outside quiet-Sun.
We note that our results (Figure 2 and top left panel of Figure 6) require cut-off frequency reduction to a value less than 2 mHz and hence
a value for $\tau_{\rm{R}}$ even smaller than 10 sec, which may be unrealistic. We, however, identify that the theoretical treatment of
the longitudinal tube mode (or the sausage mode) subject to RD by \citet{1983SoPh...87...77R} provides a formula for cut-off frequency 
that yields numbers differing from that obtained for the acoustic-gravity mode within the 2D magnetic flux tubes treated by \citet{2008ApJ...676L..85K},
and provides a better match to the observations here: under the near-isothermal photospheric conditions, Eqn. (30) of \citet{1983SoPh...87...77R}
yields for the tube cut-off frequency, $\omega_{\rm{tc}} = c_{\rm{T}}/2H_{p}$, where $H_{p}$ is the pressure scale-height, $c_{\rm{T}}$ is 
the tube speed given by $c_{\rm{T}}^{2} = c^{2}v_{\rm{A}}^{2}/(c^{2}+v_{\rm{A}}^{2})$ and $v_{\rm{A}}$ is the Alfven speed. The tube speed $c_{\rm{T}}$
in general, for photospheric conditions of $v_{\rm{A}} \approx c$, is lower than both $c$ and $v_{\rm{A}}$.
Noting that without RD the tube cut-off is almost the same as the quiet-Sun cut-off of 5.2 mHz \citep{1981A&A....98..155S,1983SoPh...87...77R}, the
above limit for tube cut-off under RD in isothermal conditions yields $\omega_{\rm{tc}} \approx $ 1.3 mHz, with $H_{p}$ = 150 km, $c_{\rm{T}}$ = 5 km s$^{-1}$
and $v_{\rm{A}}^{2}$ = $c^{2}$ = $\gamma g H_{p}$, $\gamma$ and $g$ being the adiabatic index and acceleration due to gravity. 
We also note that the early observations by \citet{1978SoPh...59...49G} has considerable
overlap with the height ranges covered in this work, and agree better with the results for $t_{\rm{ph}}$ in quiet-Sun network magnetic fields. 
In any case, it is clear from the whole scenario above that a mechanism different from 
that of magnetic inclination assisted channeling of ambient acoustic waves to generate the observed (Figure 6, top left panel) low-frequency 
(2 - 4 mHz) waves within the small-scale less-inclined or nearly vertical quiet-network magnetic fields is necessary. We come back to this aspect and discuss 
this further in relation to the vast amount of work that addresses generation of waves within flux tubes or sheaths in \S \ref{sum}.

In the plage and spot regions, as depicted in the top right sub-panels of Figures 6 and 7 and the top panels of Figure 8, 
$t_{\rm{ph}}(\theta,\nu)$ and $t_{\rm{ph}}(B_{\rm{tot}},\nu)$ show striking differences in their character as compared to the quiet-network.
Firstly, the nearly vertical ($\theta <$ 30\textdegree) core regions of these large magnetic structures show much 
reduced or no propagation signals in the 2 - 4 mHz range indicating a similarly reduced vigour or the absence of 
the wave-generation process, that we inferred for the small-scale quiet-network fields. 
Secondly, almost all of the low-frequency propagation signals show a tight correlation with the cut-off frequency
reduction according to $\nu^{B}_{\rm{ac}}=\nu_{\rm{ac}}cos\theta$: this is especially clear in the 430 km (1600~{\AA})
results in the top right sub-panels of Figures 6 and 7 for plages and the top panels of Figure 8 for the sunspot.
These dominant influences of magnetic inclination $\theta$ cause the spatial pattern
of smallest $\nu$ ($\approx$ 2 mHz) from the outer edges (larger $\theta$) to $\nu$ of 5 mHz (3 minutes) and higher 
towards the cores of plages or umbrae of spots. These results for spots agree with those of earlier authors \citep{2006ApJ...647L..77M,
2016ApJ...830L..17Z,2016NatPh..12..179J}. Further, the results here show that plage structures too exhibit similar $\theta - \nu$
patterns for propagating waves and contrast them with those of the small-scale quiet-network fields.

\subsubsection{Wave-energy Fluxes }\label{res2.f}
As to the energy fluxes $F(\theta,\nu)$ and $F(B_{\rm{tot}},\nu)$, all magnetic structures show a similar $\nu$ dependence,
which is dominantly determined by the background acoustic spectrum. Against inclination $\theta$ and strength $B_{\rm{tot}}$, the influences
of $t_{\rm{ph}}$ in $F$ are clear: in the quiet-network, $F$ has a broad peak over $\theta$ extending from 20 to 70\textdegree,
arising from the distinct wave-generation processes that we identified and discussed above in \S\ref{res2.ph} as happening within the small-scale flux 
concentrations in addition to the magnetic inclination caused reduction in $\nu_{\rm{ac}}$; otherwise, in the larger structures, 
plage and sunspot, $F$ peaks at about $\theta$ = 60\textdegree, arising primarily from the filtering-in of ambient acoustic
waves due to the reduced acoustic cut-off $\nu^{B}_{\rm{ac}}$. Theoretical studies of interaction between p-modes and magnetic fields 
identify this particular process, among several mode-conversion mechanisms \citep{2016GMS...216..489C}, as one by which the
incident acoustic waves continue along the magnetic fields as a slow acoustic mode, and has been termed as the `ramp effect'
\citep{2007AN....328..286C}: this happens when the wave-vector is largely aligned or have small angles with magnetic field direction.
Our results here, particularly the $\theta$ = 60\textdegree~peaked $F(\theta,\nu)$ shown in Figures 6 - 8, are in good agreement 
with the results from numerical simulations of mode-conversion processes in spot-like magnetic structures \citep{2013MNRAS.435.2589C}. 

Except for the sunspot (lower panels of Figure 8), $F(B_{\rm{tot}},\nu)$ peak roughly around 100 G both in the quiet and plage regions.
But, in the quiet-network, similar to the broad peak over $\theta$, the peak in $F(B_{\rm{tot}},\nu)$ does extend over to larger $B_{\rm{tot}}$
for reasons discussed above. For the sunspot, the distinct smaller amplitude peak in $F$ at $\nu$ centered around 6 mHz over the nearly
vertical ($\nu <$ 20\textdegree) and strong field ($B_{\rm{tot}} > $2500 G) umbral region is clear in the lower panels of Figure 8; these
correspond to the well-known 3-minute umbral oscillations.

\subsection{Mean Wave-energy Fluxes}\label{res3}

An important quantity that we can derive from our work here is the mean wave-energy flux over the region of observation,
especially at low (2 - 5 mHz) frequencies where the magnetic fields play key roles \citep{2006ApJ...648L.151J}, 
both in channelling such waves from the ambient medium and in generating them in-situ. As our foregoing analyses and results show with
sufficient clarity, the multi-height observables provided by the HMI and AIA (1700~{\AA} and 1600~{\AA}) instruments, along 
with the photospheric vector magnetic fields, form an excellent set of data to estimate the wave-energy supply to the chromospheric 
layers. Energy flux estimates, in our method here, depend on the atmospheric height ranges $\triangle z$ and 
$\triangle z_{\rm{v}}$ and the density scale-height $H_{\rho}$ used for calculating group velocities $v_{\rm{g}}$ and velocity power $P$ at
chromospheric heights. As detailed in \S\ref{e_est}, although the method of calculations here is entirely
diferent from that of \citet{2005Natur.435..919F}, the 5.2 - 10 mHz quiet-Sun mean wave-energy flux that we derive, for
a realistic set of values of $\triangle z = $190 km, $\triangle z_{\rm{v}} = $210 km and $H_{\rho} = $110 km corresponding to AIA 1700~{\AA}
height of 360 km, is 286.2 W m$^{-2}$ (refer to Figure 11), the same as estimated by \citet{2005Natur.435..919F} and 
\citet{2006ApJ...648L.151J}. This agreement, without any additional scaling as done by \citet{2006ApJ...648L.151J} to calibrate their estimates, 
makes our estimates for the lower frequency (2 - 5.2 mHz) energy fluxes directly comparable with that of these authors.
Again, as pointed out in \S\ref{e_est}, although the 5.2 - 10 mHz energy flux is likely an underestimate by a factor of $\approx e$ for
$\nu > \nu_{\rm{ac}}$, use of magnetically modified evolution of velocities in height $\exp{[\triangle z_{\rm{v}}/4H_{\rho}]}$, rather than 
$\exp{[\triangle z_{\rm{v}}/2H_{\rho}]}$, is important when much of the propagation at $\nu < \nu_{\rm{ac}}$ is within 
magnetic fields as observed here ({\it cf.} Figures 3 - 5). However, to the extent that there could be other physical
causes than the magnetic fields, for example, non-adiabtaicity, for wave propagation at $\nu < \nu_{\rm{ac}}$, our
estimates likely are lower limits. 

The mean energy fluxes against frequency over the quiet, plage and spot areas estimated at the two heights,
360 and 430 km, are shown in Figure 11. The results for the quiet-network in Figure 11 is for the rectangular sub-region enclosed
by the blue dotted-line boundary within the August 03, 2010 region shown in Figure 1.
At height 360 km (AIA 1700~{\AA}), using the above values for $\triangle z$, $\triangle z_{\rm{v}}$ and $H_{\rho}$ that give a 
flux of 286.2 W m$^{-2}$ over 5.2 - 10 mHz, the 2 - 5.2 mHz quiet-Sun mean energy flux is 2251.5 W m$^{-2}$. And, at height
430 km (AIA 1600~{\AA}), where we have $\triangle z = $260 km, $\triangle z_{\rm{v}} = $280 km and the same $H_{\rho} = $110 km, the
mean energy fluxes over 5.2 - 10 mHz and 2 - 5.2 mHz are 243.6 W m$^{-2}$ and 2592.0 W m$^{-2}$, respectively. 
The above 2 - 5.2 mHz quiet-Sun mean energy fluxes in the range of 2.25 - 2.6 kW m$^{-2}$ is about twice
that reported by \citet{2006ApJ...648L.151J} for the same range of heights as probed here. Our estimated range for the low-frequency
wave-energy supply facilitated by the quiet-Sun magnetic field is, thus, a much larger fraction of that required for
balancing the average quiet-chromospheric radiative losses of 4.6 kW m$^{-2}$ according to the semi-empirical VAL3-C model of
\citet{1981ApJS...45..635V}. 

As shown and discussed in \S\ref{res2.ph}, we have deduced that the operation of a mechanism different from that due to magnetic
inclination assisted propagation of low-frequency (2 - 5.2 mHz) waves within the small-scale quiet network and which produces
waves that evolve faster to steepen at the lower chromospheric heights is necessary. It is instructive to estimate the contribution by this new 
mechanism in comparison to that due to reduced cut-off frequency. In the quiet-Sun energy flux maps shown in the lower left panels of Figure 6, 
we take a ratio of flux summed over all locations where $\theta < \theta_{\rm{ac}}$ = $cos^{-1}(\nu/\nu_{\rm{ac}})$ 
to that over the whole region. The results are plotted in the right panel of Figure 9, for both the heights.
At 3 mHz and at 360 km, up to about 32\% of the energy flux over the whole region is contributed by waves, which are propagating  
wthin the small-scale magnetic fields where the inclination $\theta < \theta_{\rm{ac}}$ and hence are not the ambient acoustic waves
let through via the reduction in acoustic cut-off. Over the 2 - 5.2 mHz range, the net energy flux by such waves is about 25\%. 
At 430 km, the energy flux of $\theta < \theta_{\rm{ac}}$ waves is smaller at about 20\% of that of the whole region.

The mean wave-energy fluxes over the plage and spot regions too have been estimated and are shown in Figure 11.
The larger flux concentrations, due to larger areas of inclined fields, allow a larger flux of acoustic waves, and the numbers
are marked within the panels of Figure 11. The interesting reduction in the fluxes of higher frequency (5.2 - 10 mHz) waves
is due to the large-scale `shadows' around these large structures, clearly seen in the right panels of Figures 4 and 5, and
are associated with the magnetic canopy around them. The missing energy at chromospheric levels could perhaps be the reason for
the enhanced power of photospheric high-frequency waves called the `acoustic halos' \citep[see][and reference therein]{2013SoPh..287..107R},
which have been modeled as due to fast-wave refraction \citep{2016ApJ...817...45R}.
\begin{figure*}[ht] 
$\begin{array}{rl}
    \includegraphics[width=0.5\textwidth]{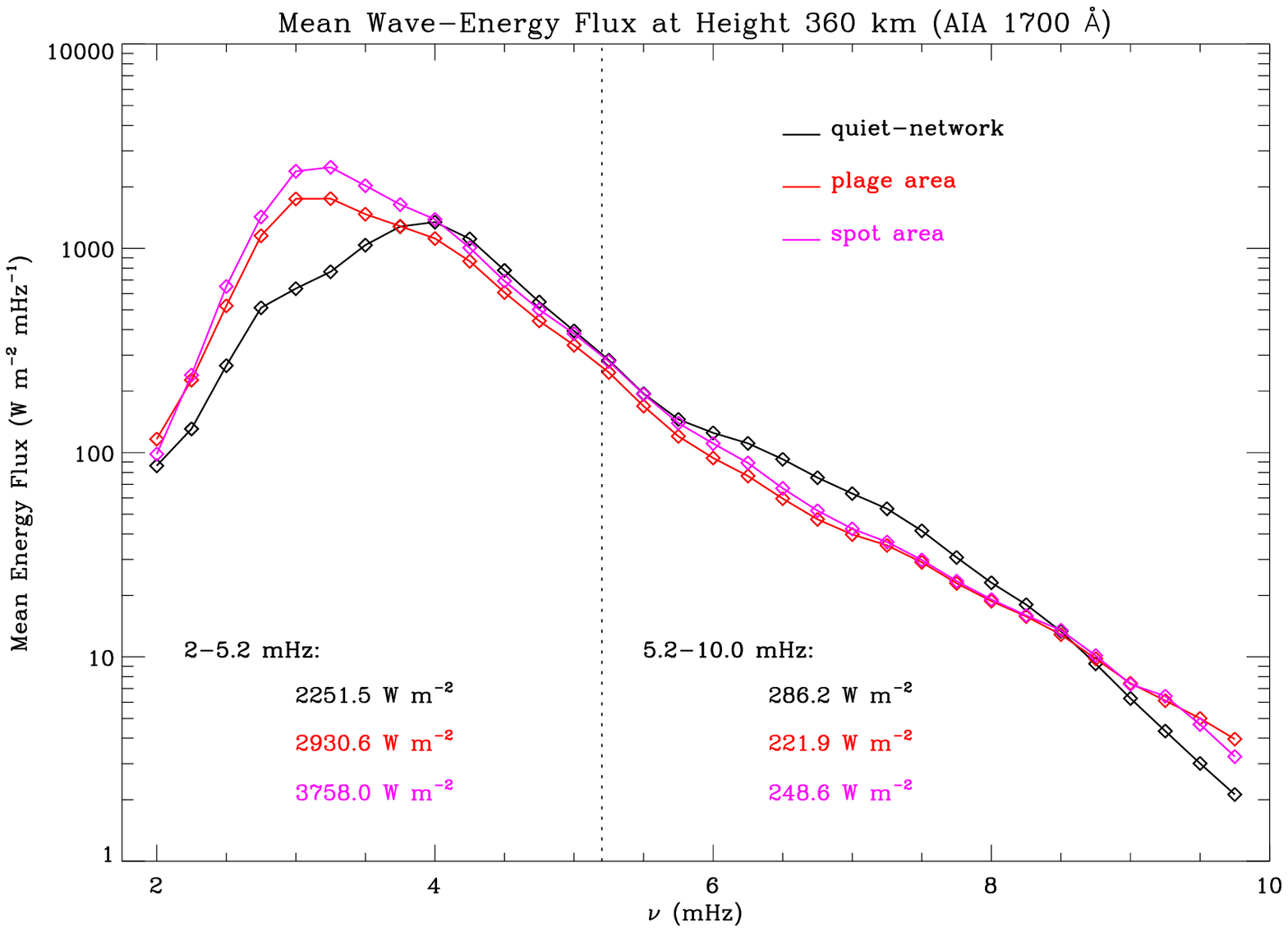} &
    \includegraphics[width=0.5\textwidth]{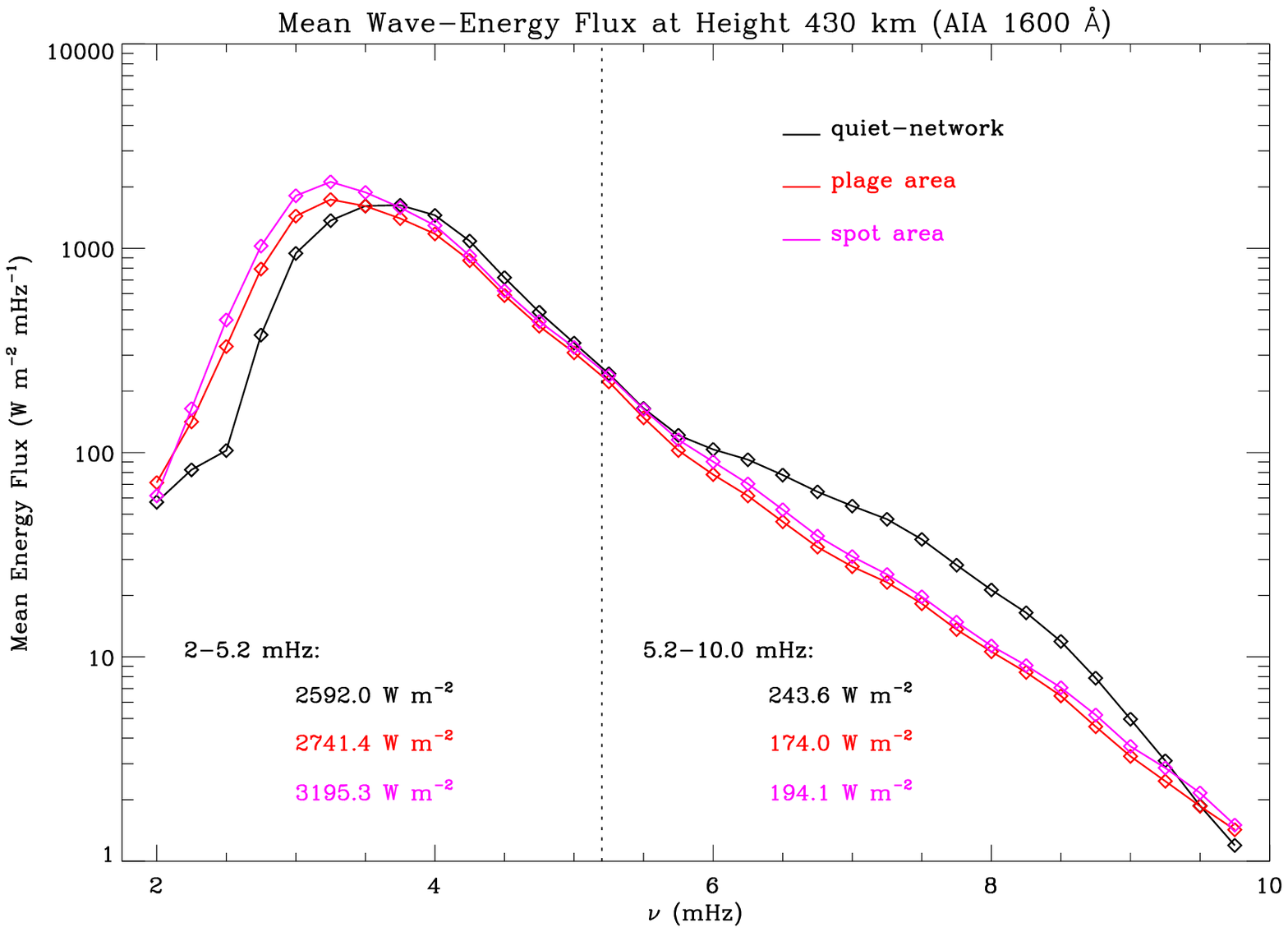}\\
\end{array}$
\caption{Mean wave-energy fluxes against frequency at heights 360 km (AIA 1700~{\AA}, left panel) and 430 km (AIA 1600~{\AA}, right panel).
        The three curves, black, red and pink correspond to mean fluxes over quiet-network, plage and spot areas, respectively.
        In each panel, we have marked the freqeuncy integrated fluxes over 2 - 5.2 mHz and 5.2 - 10 mHz ranges. The vertical dotted line
        marks the quiet-Sun acoustic cut-off frequency of 5.2 mHz. See text for full discussion.}\label{fig11}
\end{figure*}
\section{Summary and Discussion}\label{sum}
In this paper, we have studied in detail the relationships between magnetic fields and (magneto-)acoustic wave-propagation in the lower solar
atmosphere utilising the excellent quality data provided by the instruments HMI and AIA onboard SDO.
We have estimated phase travel-times of acoustic waves in the atmosphere using cross-spectra of 
intensities between two heights, used them to derive group velocities of these waves, and combined them with wave-power
derived from velocity data (HMI) to calculate energy fluxes. We have examined in detail the relations between the above
derived quantities and the vector magnetic field -- inclination and strength -- in quiet,
plage and sunspot regions. While establishing connections with and confirming several known results in the literature
as discussed in the sub-sections of \S\ref{main}, we have also derived some new results, which we summarise below.

(1) Between the upper photospheric heights of 150 - 170 km (velocities and intensities observed by SDO/HMI) and lower 
chromospheric heights of 360 - 430 km (1700~{\AA} and 1600~{\AA} intensities observed by SDO/AIA), we have identified 
and shown that the relatively less-inclined, nearly vertical magnetic field elements in the quiet-Sun channel a significant amount of 
acoustic waves of frequencies, in the range of 2 - 4 mHz, that are smaller than the theoretical minimum for the reduction 
in acoustic cut-off due to magnetic inclination (\S\ref{res2.ph}, Figures 6 and 7)\citep{1977A&A....55..239B}.\\
(2) Between the heights of 360 and 430 km, the phase travel times of the above low frequency waves within the isolated and 
nearly vertical magnetic elements do not show an increase expected over height, rather a decrease for the lowest freqeuncies (2 - 2.5 mHz),
({\it c.f.,} Figure 9, left panel), indicating wave-steepening or dissipation. \\
(3) Propagating waves observed within the larger magnetic structures, plages and a sunspot, confirm to the
reduction in cut-off due to magnetic inclination, {\it i.e.,} obey the relation $\nu^{B}_{\rm{ac}}=\nu_{\rm{ac}}cos\theta$ for
the minimum $\nu$ of propagation \citep{1977A&A....55..239B,2006ApJ...647L..77M}, and do not exhibit the 
characteristics described in (1) and (2) above. Instead, phase travel times of waves propagating along the inclined
field do increase through the height ranges probed indicating propagation beyond 430 km.\\
(4) The energy flux of acoustic waves within magnetic structures, small and large, peak around magnetic inclination
$\theta=$60\textdegree~(bottom panels, Figures 6 - 8), except that the small-scale quiet-Sun magnetic fields exhibit a 
broader peak extending down to $\theta \approx$20\textdegree~(bootom left panel, Figure 6). This result for the larger structures
provides an observational verification of the numerical simulations of mode-conversion process, 
especially the ramp effect \citep{2013MNRAS.435.2589C}.\\
(5) The mean energy flux over the quiet-Sun (or quiet magnetic network), over the 2 - 5 mHz frequency region, between the upper 
photosphere and heights corresponding to AIA 1700~{\AA} and 1600~{\AA} intensities is in the range of 2.25 to 2.6 kW m$^{-2}$ 
(\S\ref{res3}, Figure 11), which is about twice the previous estimates \citep{2006ApJ...648L.151J}. Out of this total quiet-Sun
acoustic flux, about 25 to 30\% is due to the newly identified wave-propagation process summarised in results (1) and (2) above
(right panel, Figure 9).

Since the waves of frequencies 2 - 4 mHz found in result (1) are lower than the theoretical minimum given by $\nu^{B}_{\rm{ac}}=\nu_{\rm{ac}}cos\theta$
they certainly are not the ambient acoustic waves propagating because of magnetic-inclination-assisted cut-off frequency reduction. 
We, in \S\ref{res1}, ruled out artifacts or contamination due to height-dependent RD as the cause of result (1). And, 
in \S\ref{res2.ph}, we discussed the connections between our result (1) and the observations and models based on RD by
\citet{2006ASPC..358..465C} and \citet{2008ApJ...676L..85K} on the one hand and those of \citet{1978SoPh...59...49G} and \citet{1983SoPh...87...77R}
on the other. Both the above earlier observational results show increasing $t_{\rm{ph}}$ over heights that extend well above the range covered here 
and hence are in contrast with our result (2), {\it viz.} the low-frequency (2 - 4 mHz) waves within the small-scale near-vertical magnetic fields 
evolve over height faster ({\it i.e.,} shorter  $t_{\rm{ph}}$) than the ambient acoustic waves do along inclined magnetic fields -- this particular
property indeed is seen in some wave simulation studies and we come back to it in the next paragraph.
We note that the above propagating waves within the near-vertical fields are accompanied by the slightly larger, but still well below the 
quiet-Sun acoustic cut-off, frequency waves in the inclined boundaries of the same small-scale magnetic structures with their $t_{\rm{ph}}$ 
(Figure 6, top right panel for quiet-Sun) increasing over height. It should be pointed out that the above earlier authors did not have magnetic 
inclination information in their observations and hence it was not clear that the increasing $t_{\rm{ph}}$ signals at the low frequencies that 
they measure at higher in the chromosphere were still associated with that of RD origin in photospheric vertical magnetic field; at heights above 
500 km, vertical flux tubes must have expanded to fill the chromospheric volume and hence waves let out due to inclined magnetic field might spatially 
overlap with those propagating due to RD and in addition, mode conversion between fast modes to slow waves or between kink modes to slow modes will 
confuse the identification of the origin of propagating waves. 
Hence, the scenario presented by our results (1) and (2) together appear distinct from that of above authors' observational results. 
However, as discussed in detail in \S\ref{res2.ph}, we do find that the case of the tube mode (or sausage mode) subject 
to strong photospheric RD as theorised by \citet{1983SoPh...87...77R} provides an interesting explanation for result (1), although a detailed 
study of non-linear development and shock-formation of the tube mode subject to RD is necessary to explain results (1) and (2) together.
Further, for flux elements of small horizontal size, a careful treatment of horizontal radiative exchange is important while treating the
RD due to vertical radiative losses, because such multi-dimensional radiative transfer can vastly modify the wave motions \citep{1987A&A...183...91K}, 
{\em e.g.} making the waves overstable (countering the RD) \citep{1966ApJ...143..871M,1989A&A...209..399M,2000ApJ...544..522R}.

An alternative possibility is the in-situ wave generation process, 
accompanied by changes in the atmospheric stratification within the magnetic elements so as to allow propgation 
at frequencies well below the acoustic cut-off.
Moreover, since result (3) shows that such low-frequency propagation is not seen within the cores (or less inclined central regions) of larger
structures, plages and sunspot, the above in-situ process must be related to the flux elements being small (horizontally).
The long-period network oscillations within the quiet magnetic-network as originally observed by
\citet{1993ApJ...414..345L} have peak frequencies around 2.5 mHz ($\approx$ 7 minutes) coinciding with our finding of dominant propagating
waves in this same frequency range (Figure 6, {\it top left panel}). There have been a lot of work, in the literature, that 
addresses generation of waves within flux tubes or sheaths due to the buffetting action of external convection or waves 
\citep{1993SoPh..143...49C,1995ApJ...448..865M,1999ApJ...519..899H,2003ApJ...599..626B,2003ApJ...585.1138H,2008ApJ...680.1542H,2009A&A...508..951V} 
or in-situ generation within them by magneto-convective processes \citep{2011ApJ...730L..24K,2016ApJ...827....7K}. 
The dominant periods of longitudinal (acoustic) waves generated within the flux tubes or sheaths in the above two
distinct mechanisms are quite different: the former one utilising an external driving, either by transverse motions 
imparted by granules or through coupling with the ambient p modes, predict longitudinal waves of frequencies larger 
or at the tube cut-off frequency \citep{1981A&A....98..155S,1999ApJ...519..899H}, which is nearly the same as quiet-Sun 
acoustic cut-off of $\approx$ 5.2 mHz (3 minutes). However, the latter magneto-convective process within flux tubes or sheaths
drive upward propagating longitudinal waves over a broad frequency range, 1 - 10 mHz, with peak power around 4 mHz 
\citep{2016ApJ...827....7K}, providing possible explanations for our observational results in Figures 6 and 7 for 
the quiet-network. We do note that \citet{2016ApJ...827....7K} indeed invoke possible contributions of magnetic inclination 
for the existence of low-frequency ($< $5.2 mHz) oscillations while cautioning that the simple estimates of acoustic 
cut-off frequency \citep{1981A&A....98..155S} do not hold good in the non-linear wave evolution
typical in the atmospheric layers. Further, although the transverse motions of flux tubes or sheaths excite
longitudinal tube modes that propagate at or above the cut-off frequency (5 mHz), \citet{2008ApJ...680.1542H} do point out the
possibility of the kink mode of much lower cut-off frequency (around 3.3 mHz) mode-converting to cause longitudinal waves
at chromospheric heights. This again cannot explain our observations as we have longitudinal waves propagating from photospheric
heights at frequencies well below 3.3 mHz. We however note that the simulations of \citet{2008ApJ...680.1542H} show an important 
difference between longitudinal waves excited within flux tubes by transverse motions and those, from the ambient acoustic
waves, propagating along inclined field lines: the former evolve quickly to become shocks at chromospheric heights whereas the
latter propagate through to higher layers. Our result (2) confirms this behaviour, although the observed longitudinal waves
are well below the cut-off frequency. 

In summary, the RD mechanism acting on the tube or sausage mode \citep{1983SoPh...87...77R} or the one due to
magneto-convective processes \citep{2011ApJ...730L..24K,2016ApJ...827....7K} discussed in the above two paragraphs, respectively,
are relevant for the observations here. We, however, caution that more detailed examination, with multi-height velocity and
intensity data, of our observational findings as well as the above theoretical and numerical models is necessary to pinpoint 
the exact mechanisms behind the low-frequency (lower than the theoretical miniumum allowed by magnetic inclination) lonigitudinal 
propagating waves in the small-scale magnetic fields (Figure 6).
As regards the role of small-scale magnetic fields, which \citet{2006ApJ...648L.151J} called as magneto-acoustic portals,
our result (5) shows that about half of 4.6 kW m$^{-2}$ \citep{1981ApJS...45..635V} required to balance the 
average quiet-chromospheric radiative losses is carried by the energetic low frequency (2 - 5.2 mHz) acoustic waves harbored by them.
Our results thus further corraborate the role of small-scale magnetic fields in chromospheric heating and in explaining the basal 
flux of emissions \citep{1989ApJ...341.1035S,2005Natur.435..919F}.

Extensions of the study reported here would benefit a lot from a much finer height coverage of the photosphere - chromosphere
region. Currently being built Solar Ultraviolet Imaging Telescope \citep[SUIT][]{2016SPIE.9905E..03G,Durgesh_suit} for the
upcoming Aditya-L1 mission of the Indian Space Research Organisation (ISRO) will have such capabilities: 
imaging the Sun within the wavelength band of 2000{--}4000~{\AA} covered by 11 different filters providing near-simultaneous 
observations.

\begin{acknowledgements}
C.R.S. and D.T. acknowledge support from the Max-Planck India Partner Group of MPS on ``Coupling and Dynamics of the Solar Atmosphere'' 
at IUCAA. The partner group is funded by MPG and DST(IGSTC), Gov. of India. Data-intensive numerical computations required in this 
work were carried out using the High Performance Computing facility of the Indian Institute of Astrophysics, Bangalore. This work has 
utilised extensively the HMI/SDO and AIA/SDO data pipeline at the Joint Science Operations Center (JSOC), Stanford University. We
also thank an anonymous referee of this paper for prompting us to address the effects of radiative damping.

\end{acknowledgements}

\end{document}